\documentclass[12pt]{iopart}

\usepackage{graphicx}
\usepackage{multicol}
\usepackage{dcolumn}
\usepackage{bm}
\usepackage{hyperref}
\usepackage{booktabs}
\usepackage{xcolor}
\usepackage[utf8]{inputenc}
\usepackage{caption}
\usepackage{subcaption}
\usepackage{soul}

\definecolor{byzantium}{HTML}{702963}
\newcommand\nmesh{\texttt{Nmesh}}

\begin{document}

\title{The new discontinuous Galerkin methods based numerical relativity
program \nmesh}


\author{Wolfgang Tichy${}^1$, Liwei Ji${}^2$, Ananya Adhikari${}^1$,
Alireza Rashti${}^{3,4}$, Michal Pirog${}^1$}
\address{${}^1$ Department of Physics, Florida Atlantic University,
              Boca Raton, FL 33431, USA}
\address{${}^2$ Rochester Institute of Technology,
              Rochester, NY 14623, USA}
\address{${}^3$ Institute for Gravitation \& the Cosmos, 
         The Pennsylvania State University, University Park PA 16802, USA}
\address{${}^4$ Department of Physics, The Pennsylvania State University, 
                  University Park PA 16802, USA}



%
\newcommand\be{\begin{equation}}
\newcommand\ba{\begin{eqnarray}}

\newcommand\ee{\end{equation}}
\newcommand\ea{\end{eqnarray}}
\newcommand\p{{\partial}}
\newcommand\remove{{{\bf{THIS FIG. OR EQN. COULD BE REMOVED}}}}

%

\begin{abstract}

Interpreting gravitational wave observations and understanding the physics
of astrophysical compact objects such as black holes or neutron stars
requires accurate theoretical models. Here, we present a new numerical
relativity computer program, called \nmesh, that has the design goal to
become a next generation program for the simulation of challenging
relativistic astrophysics problems such as binary black hole or neutron star
mergers. In order to efficiently run on large supercomputers, \nmesh~uses a
discontinuous Galerkin method together with a domain decomposition and
mesh refinement that parallelizes and scales well. In this work, we
discuss the various numerical methods we use. We also present results of
test problems such as the evolution of scalar waves, single black holes and
neutron stars, as well as shock tubes. In addition, we introduce a new
positivity limiter that allows us to stably evolve single neutron stars
without an additional artificial atmosphere, or other more traditional
limiters.

\end{abstract}

\maketitle

\section{Introduction}

In August 2017, a binary neutron star merger has been observed
by detecting its gravitational wave
signal~\cite{TheLIGOScientific:2017qsa} together with an
electromagnetic counterpart (across the whole electromagnetic
spectrum)~\cite{LIGOScientific:2017zic,Coulter:2017wya}.
This and similar observations~\cite{LIGOScientific:2018mvr,
LIGOScientific:2020ibl,LIGOScientific:2021qlt} have started
a new era of multi-messenger astronomy~\cite{GBM:2017lvd} and
have opened a new window to the Universe, that allows us to
measure and understand phenomena
related to the equation of state~(EoS) at supranuclear densities, the
production of heavy elements via rapid neutron capture~(r-process)
nucleosynthesis, and cosmological
constants~\cite{GBM:2017lvd,Abbott:2017xzu,Radice:2017lry,Most:2018hfd,
Metzger:2019zeh,Dietrich:2020efo}.

Accurate theoretical models are required for creating gravitational wave and
electromagnetic templates to interpret the observations, and to extract
all the information contained in such signals about the properties of the
binary.
While there are analytical models to describe compact object coalescence,
as long as the objects are well separated~\cite{Blanchet:2013haa},
the highly non-linear regime around the moment of merger is
only accessible through simulations employing
full numerical relativity (NR).
To carry out such simulations, various computer programs have been developed,
e.g., BAM~\cite{Bruegmann:2006ulg,Thierfelder:2011yi,Dietrich:2015iva},
Einstein Toolkit~\cite{Loeffler2012,EinsteinToolkit:2020_11},
NRPy+~\cite{Ruchlin:2017com},
SACRA-MPI~\cite{Kiuchi:2019kzt}, and
SpEC~\cite{SpECWebsite,Boyle:2019kee}.
Given that current detectors have a relatively high noise level,
the numerical errors in these computer programs are not the main
limiting factor when comparing observations and simulations.

However, the arrival of a new generation of detectors in the near future,
like Cosmic Explorer~\cite{Reitze:2019iox},
the DECi-hertz Interferometer Gravitational-wave Observatory
(DECIGO)~\cite{Kawamura:2020pcg},
Einstein Telescope~\cite{Punturo:2010zz},
LIGO Voyager~\cite{Adhikari:2019zpy},
the Laser Interferometer Space Antenna (LISA)~\cite{LISA:2017pwj},
NEMO~\cite{Ackley:2020atn}, and
TianQin~\cite{TianQin:2015yph},
will allow for observations with much higher signal-to-noise ratios.
Therefore, in order to not bias the interpretation of the observed data,
future NR computer programs will be required to better model micro-physics
and also to deliver simulations with a higher accuracy. In principle higher
accuracy can be achieved by current computer programs by simply increasing
the resolution, while at the same time using more computational cores. This,
however, will likely raise the computational cost to a level that is no
longer affordable, because conventional NR computer programs do not scale
well enough when we want to use hundreds of thousands of computational
cores.

Consequently, a new campaign in the NR community is taking place
to upgrade and develop computer programs that scale well and thus
have a chance to achieve the accuracy needed for future observations.
Examples of such next generation programs are
BAMPS~\cite{Bugner:2015gqa,Hilditch:2015aba},
GRaM-X~\cite{Shankar:2022ful,CarpetX_web},
Dendro-GR~\cite{Fernando:2018mov},
ExaHyPE~\cite{Koppel:2017kiz}
GR-Athena++~\cite{Daszuta:2021ecf}
GRChombo~\cite{Clough:2015sqa,Andrade:2021rbd}
SpECTRE~\cite{Kidder:2016hev,Deppe:2021bhi}, and
SPHINCS\_BSSN~\cite{Rosswog:2020kwm}.

In this work, we present a new computer program, called \nmesh, that
aims to be one of these next generation programs.
One of the main features of \nmesh~is its use of
discontinuous Galerkin (DG) methods.
The DG method for hyperbolic conservation laws has been
introduced in~\cite{Cockburn:1991a,Cockburn:1989a,
Cockburn:1989b,Cockburn:1990a,Cockburn1998a}. Only
in recent years it has been explored by the numerical relativity community
to evolve the Einstein equations and the equations of general relativistic
hydrodynamics~\cite{Radice:2011qr,Bugner:2015gqa,Teukolsky:2015ega,
Kidder:2016hev,Dumbser:2017okk,Fambri:2018udk,Cao:2018myc,Deppe:2021ada,
Deppe:2021bhi}. It has two main advantages when compared to more
traditional finite difference or finite volume methods. First, when the
evolved fields are smooth, a DG method can be exponentially convergent and
thus much more efficient than traditional methods. Second, because of
the way boundary conditions between adjacent domains are
imposed within a DG method, there is less communication overhead
when domains are distributed across many computational cores. This
is expected to result in better scalability in a future where many
groups will want to use hundreds of thousands or possibly even
millions of computational cores.

The main purpose of this paper is to describe and test \nmesh. As will be
discussed below, the main novelties when compared to other programs such as
SpECTRE (as described in~\cite{Deppe:2021bhi})
are a simplified treatment of the normal vectors and induced metric
on domain boundaries, as well as certain positivity limiters that allow us
to evolve single neutron stars without any additional limiters.

In Sec.~\ref{num_methods} we describe the DG method and other numerical
methods we use. This is followed by a discussion of the effectiveness
of our parallelization in Sec.~\ref{parallelization_strategy}.
In Sec.~\ref{evo_system_tests} we test how well \nmesh~can evolve
various systems such as scalar waves, black holes, neutron stars, as well
as some shock waves. We summarize and discuss our results in
Sec.~\ref{discussion}.
Throughout the article, we use geometric units, in which $G=c=1$, as well as
$M_{\odot}=1$.
Indices from the middle of the Latin alphabet, such as $i$, run from 1 to 3
and denote spatial indices, while indices from the beginning of the Latin
alphabet, like $a$, and also Greek indices, such as $\mu$, run from 0 to 3 and
denote spacetime indices.


\section{Numerical methods}
\label{num_methods}

In this section, we present the various numerical methods we use to perform
simulations with hyperbolic evolution equations.

\subsection{The discontinuous Galerkin method}

In \nmesh, we use a discontinuous Galerkin (DG) method to discretize
evolution equations. Often, these evolution equations come from general
relativistic conservation laws of the form
\be
\label{Cons0}
\nabla_{\mu} J^{\mu} = S ,
\ee
where $S$ is a possible source term. The covariant divergence on the left
hand side can be written in terms of coordinate derivative using
$\nabla_{\mu} J^{\mu} =
\frac{1}{\sqrt{|g|}}\partial_{\mu} (\sqrt{|g|} J^{\mu})$, where
$g$ is the determinant of the 4-metric $g_{\mu\nu}$.
In terms of the standard 3+1 decomposition~\cite{Arnowitt62}, the
4-metric is written
as
\be
\label{4metric}
ds^2 = g_{\mu\nu}dx^{\mu}dx^{\nu}
= -\alpha^2 dt^2 + \gamma_{ij}(dx^i + \beta^i dt)(dx^j + \beta^j dt) ,
\ee
where $\gamma_{ij}$ is the spatial metric on $t=const$ slices, and $\alpha$
and $\beta^i$ are called lapse and shift. We can show that
$\sqrt{|g|} = \alpha\sqrt{\gamma}$, where $\gamma$ is the determinant
of the 3-metric $\gamma_{ij}$. Thus, Eq.~(\ref{Cons0}) is equivalent to
\be
\label{Cons1}
\partial_t (\sqrt{\gamma}\alpha J^t) + \partial_i (\sqrt{\gamma}\alpha J^i)
= \sqrt{\gamma}\alpha S .
\ee
Usually, we introduce the new variables
\be
u = \sqrt{\gamma}\alpha J^t, \ \ \
f^i =  \sqrt{\gamma}\alpha J^i, \ \ \
s = \sqrt{\gamma}\alpha S ,
\ee
so that Eq.~(\ref{Cons1}) finally yields
\be
\label{Cons2}
\partial_t u +  \partial_i f^i = s .
\ee
Note that the flux vector $f^i$ is usually a function $f^i(u)$, that depends
on $u$. For brevity, we omit this dependence in most equations.

To discretize the spatial derivatives in Eq.~(\ref{Cons2}), we first integrate
against a test function $\psi$ with respect to the coordinate volume $d^3x$
over a certain region $\Omega$. We obtain
\be
\label{Integ0}
\int (\psi\partial_t u + \psi\partial_i f^i) d^3x = \int \psi s d^3x .
\ee
The second term is now integrated by parts using
\be
\label{int_by_parts}
\int \psi\partial_i f^i d^3x =
\oint \psi f^i n_i d^2\Sigma - \int f^i \partial_i\psi d^3x ,
\ee
where $d^2\Sigma$ is the surface element for integrating over the
boundary $\partial\Omega$, and $n_i$ is normal to $\partial\Omega$.
In the surface integral, the flux $f^i n_i$ at the boundary appears.
To incorporate numerical boundary conditions, $f^i n_i$
in the surface integral over the boundary is replaced by the
so-called numerical flux $(f^i n_i)^*$ (See Sec.~\ref{numflux} below).
It contains any information we need from the
other side of the boundary (such as incoming characteristic modes).
The replacement of $f^i n_i$ by $(f^i n_i)^*$ in the surface integral yields
\ba
\label{introduce_numflux}
\int \psi\partial_i f^i d^3x
& \rightarrow &
\oint \psi (f^i n_i)^* d^2\Sigma - \int f^i \partial_i\psi d^3x \nonumber \\
& = &
\oint \psi [(f^i n_i)^* - f^i n_i] d^2\Sigma + \int \psi\partial_i f^i d^3x ,
\ \ \
\ea
where in the last step we have used integration by parts again to eliminate
derivatives of $\psi$. With this replacement, Eq.~(\ref{Integ0}) becomes
\be
\label{Integ1}
\int (\psi\partial_t u + \psi\partial_i f^i) d^3x
= \int \psi s d^3x - \oint \psi [(f^i n_i)^* - f^i n_i] d^2\Sigma .
\ee

We now introduce a coordinate transformation
\be
\label{x_of_Xbar}
x^i = x^i(x^{\bar{i}}),
\ee
such that the volume we integrate over extends from $-1$ to $+1$ for all three
$x^{\bar{i}}$ coordinates. In these coordinates, Eq.~(\ref{Integ1})
reads
\be
\label{Integ2}
\int \psi \left(
\partial_t u + \frac{\partial x^{\bar{i}}}{\partial x^i} \partial_{\bar{i}} f^i
\right) J d^3\bar{x}
= \int \psi s J d^3\bar{x}
  -\oint \psi F d\bar{A} ,
\ee
where we have defined
\be
\label{def_F}
F := (f^i n_i)^* - f^i n_i ,
\ee
\be
J = \left|\det\left(\frac{\partial x^i}{\partial x^{\bar{i}}}\right) \right|,
\ee
where $J$ is called the Jacobian, and $d\bar{A}$ is the surface element on
one of the six surfaces $x^{\bar{i}}=\pm 1$, but now expressed in
$x^{\bar{i}}$ coordinates. For example on $x^{\bar{3}}=-1$,
\be
\label{dbarA_example}
d\bar{A} = \sqrt{^{(2)}\bar{\gamma}} dx^{\bar{1}} dx^{\bar{2}},
\ee
where $^{(2)}\bar{\gamma}$ is the determinant of the 2-metric induced on this
coordinate surface by the flat 3-metric $\delta_{ij}$.
The flat $\delta_{ij}$ results from our choice to integrate over the
coordinate volume $d^3x$ in Eq.~(\ref{Integ0}), without including
$\sqrt{\gamma}$. Below these $x^i$ coordinates will be chosen to be
Cartesian-like, so that they can cover all numerical domains.

Next, we expand both $u$ and $f^i$ in terms of Lagrange's characteristic
polynomials
\be
l_q(\bar{x})
= \prod_{r=0, r \neq q}^N \frac{\bar{x}-\bar{x}_r}{\bar{x}_q-\bar{x}_r}
\ee
so that, e.g.,
\be
u(\bar{x})
= \sum_{r_1=0}^N\sum_{r_2=0}^N\sum_{r_3=0}^N u_{r_1 r_2 r_3}
   l_{r_1}(x^{\bar{1}}) l_{r_2}(x^{\bar{2}}) l_{r_3}(x^{\bar{3}}) .
\ee
The $\bar{x}_r$ are $N+1$ grid points that we choose in the interval
$[-1,1]$.
For the test function $\psi$, we use these same basis polynomials, i.e.,
\be
\psi = l_{q_1}(x^{\bar{1}}) l_{q_2}(x^{\bar{2}}) l_{q_3}(x^{\bar{3}}) .
\ee

The final step is to approximate all integrals in Eq.~(\ref{Integ2})
using Gau{\ss}ian quadrature (specifically Lobatto's Integration formula
on p.~888 of~\cite{AbramowitzStegun72}),
which, in one dimension, is given by
\be
\int_{-1}^1 d\bar{x}\ g(\bar{x}) \approx
\sum_{q=0}^N w_q g(\bar{x}_q) .
\ee
Here we use the $N+1$ Legendre Gau\ss-Lobatto grid points $\bar{x}_q$, that
are defined as the extrema of the standard Legendre polynomial
$P_N(\bar{x})$ in $\bar{x}\in [-1,1]$.
The integration weights are then given by~\cite{AbramowitzStegun72}
\be
w_q = \frac{2}{N(N+1) P_N(\bar{x}_q)^2} .
\ee
The integrals in Eq.~(\ref{Integ2}) then turn into sums over products of
$l_{p}(\bar{x})$, or products of $l_{p}(\bar{x})$ and its derivative.
If we define
\be
(u,v) := \sum_{r=0}^N w_r u(\bar{x}_r) v(\bar{x}_r),
\ee
the products we encounter are $(l_q,l_r)$ and
$(l_q,\partial_{\bar{x}} l_r)$.
Since
\be
l_q(\bar{x}_r) = \delta_{qr},
\ee
we find
\be
(l_q,l_r) = w_q \delta_{qr}
\ee
and
\be
(l_q,\partial_{\bar{x}} l_r) = w_q \partial_{\bar{x}} l_r(\bar{x}_q)
=: w_q D_{qr}^{\bar{x}} .
\ee
The differentiation matrix is given by
\be
D_{qr}^{\bar{x}} = \frac{c_r}{c_q}\frac{1}{\bar{x}_q - \bar{x}_r},
\ee
where the Lagrange interpolation weights are
\be
c_q = \left[\prod_{s=0, s\neq q}^N(\bar{x}_q - \bar{x}_s)\right]^{-1} .
\ee

Putting all this together Eq.~(\ref{Integ2}) becomes
\ba
\label{Integ3}
w_{q_1} w_{q_2} w_{q_3} J_{q_1 q_2 q_3} \Big(
    \partial_t u_{q_1 q_2 q_3}
  + \frac{\partial x^{\bar{1}}}{\partial x^i}
    \sum_{r=0}^N D_{q_1 r}^{\bar{1}} f^i_{r q_2 q_3} && \nonumber\\
  + \frac{\partial x^{\bar{2}}}{\partial x^i}
    \sum_{r=0}^N D_{q_2 r}^{\bar{2}} f^i_{q_1 r q_3}
  + \frac{\partial x^{\bar{3}}}{\partial x^i}
    \sum_{r=0}^N D_{q_3 r}^{\bar{3}} f^i_{q_1 q_2 r} \Big) && \nonumber\\
= w_{q_1} w_{q_2} w_{q_3} J_{q_1 q_2 q_3} s_{q_1 q_2 q_3} && \nonumber\\
 -w_{q_2} w_{q_3} F_{q_1 q_2 q_3} \sqrt{^{(2)}\bar{\gamma}_{q_1 q_2 q_3}}
  (\delta_{q_1 0} + \delta_{q_1 N}) && \nonumber\\
 -w_{q_1} w_{q_3} F_{q_1 q_2 q_3} \sqrt{^{(2)}\bar{\gamma}_{q_1 q_2 q_3}}
  (\delta_{q_2 0} + \delta_{q_2 N}) && \nonumber\\
 -w_{q_1} w_{q_2} F_{q_1 q_2 q_3} \sqrt{^{(2)}\bar{\gamma}_{q_1 q_2 q_3}}
  (\delta_{q_3 0} + \delta_{q_3 N}),
\ea
where the Kronecker deltas on the right hand side come from
$l_q(+1) = \delta_{q N}$ and $l_q(-1) = \delta_{q 0}$.
We now divide Eq.~(\ref{Integ3}) by
$w_{q_1} w_{q_2} w_{q_3} J_{q_1 q_2 q_3}$ and use
the fact that
\be
\label{sqrtdiag_gammaii_flat}
\sqrt{^{(2)}\bar{\gamma}}/J = \sqrt{\bar{\gamma}^{\bar{i}\bar{i}}}
\ee
holds on the surface $x^{\bar{i}} = const$.
Here $\bar{\gamma}^{\bar{i}\bar{i}}$ is a diagonal component of the
inverse 3-metric obtained by transforming the flat 3-metric
$\delta_{ij}$ from $x^{i}$-coordinates to $x^{\bar{i}}$-coordinates.
This results in
\ba
\label{Integ4}
    \partial_t u_{q_1 q_2 q_3}
  + \sum_{r=0}^N \Big(
    \frac{\partial x^{\bar{1}}}{\partial x^i}
    D_{q_1 r}^{\bar{1}} f^i_{r q_2 q_3}
  + \frac{\partial x^{\bar{2}}}{\partial x^i}
    D_{q_2 r}^{\bar{2}} f^i_{q_1 r q_3}          
  + \frac{\partial x^{\bar{3}}}{\partial x^i}
    D_{q_3 r}^{\bar{3}} f^i_{q_1 q_2 r}
  \Big) && \nonumber\\
= s_{q_1 q_2 q_3}
 -\frac{\sqrt{\bar{\gamma}^{\bar{1}\bar{1}}_{q_1 q_2 q_3}}}{w_{q_1}}
  F_{q_1 q_2 q_3} (\delta_{q_1 0} + \delta_{q_1 N}) && \nonumber\\
 -\frac{\sqrt{\bar{\gamma}^{\bar{2}\bar{2}}_{q_1 q_2 q_3}}}{w_{q_2}}
  F_{q_1 q_2 q_3} (\delta_{q_2 0} + \delta_{q_2 N}) && \nonumber\\
 -\frac{\sqrt{\bar{\gamma}^{\bar{3}\bar{3}}_{q_1 q_2 q_3}}}{w_{q_3}}
  F_{q_1 q_2 q_3} (\delta_{q_3 0} + \delta_{q_3 N}) ,
\ea
which is the version we use in \nmesh's DG method.
Notice that the derivation of this DG method has mostly followed the one
introduced by Teukolsky in~\cite{Teukolsky:2015ega,Hebert:2018xbk}
and tested extensively in~\cite{Deppe:2021bhi}, except for one important
difference. Since we integrate over $d^3x$ without including the
determinant of the physical metric, we use the flat metric $\delta_{ij}$
when we construct $\bar{\gamma}^{\bar{i}\bar{i}}$
or when we normalize the normal vector $n_i$.
This same $n_i$ also enters the calculation of the fluxes and
eigenvalues discussed in subsection~\ref{numflux}.
Recall that $n_i$ arose in Eq.~(\ref{int_by_parts}) after using Gau{\ss}'s
theorem. As shown in \ref{Gauss_theorem}, Gau{\ss}'s theorem can be used
with any metric, as long as we normalize $n_i$ with this same metric.
The advantage of $\delta_{ij}$ is that it is constant and thus cannot have
any discontinuities. Our approach simplifies
the formalism as one does not have to worry about possible discontinuities
in the physical metric or the normal vector across domain boundaries,
which is an issue in the other approach~\cite{Deppe:2021bhi}. In
\ref{normalization-example} we discuss the difference between both
approaches for the well understood case of an advection equation. We find
that our approach together with the flux of Eq.~(\ref{LLF_flux}) yields the
correct upwind result, while the approach in~\cite{Deppe:2021bhi} does not,
if the physical metric is discontinuous across the boundary.
However, the true physical solution of the Einstein equations is a
continuous metric, even in the presence of shocks in the matter. Thus we
expect that both approaches will converge to the same result, because any
discontinuities in the metric should converge to zero.
We have also tried both approaches by evolving a black hole, where the
physical metric is far from flat, and where discontinuities in the physical
metric arise due to numerical errors. We find no important differences in
the numerical solution or its rate of convergence.
The discontinuities in the physical metric for this black hole case are
described in Sec.~\ref{GHG_runs} and shown in
Fig.~\ref{KS_gxx_discont_zoom}.

Let us also consider the field $u$ from Eq.~(\ref{Cons2}) for the case where
$s=0$. Then $\int u d^3x$ is exactly conserved. In formulations that
directly use Eq.~(\ref{Cons1}), and do not include $\sqrt{\gamma}$ in the
field definition, the conserved quantity is $\int U \sqrt{\gamma}d^3x$,
where $U=\alpha J^t$. In fact, for the exact solution of Eq.~(\ref{Cons2})
both integrals are identical. However, once $u$ and $U$ are expressed in
terms of basis functions (or equivalently in terms of values at grid
points), the numerical (Gau{\ss}ian quadrature) integrals over $u$ and $U$
will yield answers that differ at the level of the numerical truncation
error. Nevertheless, a correct DG formulation will preserve these numerical
integrals. Furthermore, any differences between the two numerical integrals
will converge away with increasing resolution. Thus at any finite resolution
the two approaches conserve different quantities, but in the continuum limit
both converge to the same result.

\subsection{Evolution equations in non-conservative form}

So far, we have only considered evolution equations that can be written in
conservative form as in Eq.~(\ref{Cons2}), i.e., in terms of a flux
vector $f^i$.
However, the equations describing general relativistic gravity are often
not available in this form. Rather they take the form
\be
\label{NonCons}
\partial_t u + A^i(u) \partial_i u = s .
\ee
Note that here the matrix $A^i(u)$ depends on $u$ itself.
In this case, we can still integrate against a test function $\psi$, as
before. The crucial introduction of a numerical flux
in Eq.~(\ref{introduce_numflux}) now takes the form
\ba
\label{introduce_numflux2}
\int \psi A^i\partial_i u d^3x
& \rightarrow &
\oint \psi (n_i A^i u)^* d^2\Sigma - \int u \partial_i(A^i\psi) d^3x
\nonumber \\
& = &
\oint \psi [(n_i A^i u)^* - n_i A^i u] d^2\Sigma
 + \int \psi A^i\partial_i u d^3x .
\ea
Thus, the surface integral has almost the same form, with
$(n_i A^i u)^*$ playing the role of the numerical flux.
If we again expand in Lagrange’s
characteristic polynomials, and retrace our previous steps, we find
the equivalent of Eq.~(\ref{Integ4}). We obtain
\ba
\label{IntegA}
    \partial_t u_{q_1 q_2 q_3}
  + \sum_{r=0}^N A^i_{q_1 q_2 q_3} \Big(
    \frac{\partial x^{\bar{1}}}{\partial x^i}
    D_{q_1 r}^{\bar{1}} u_{r q_2 q_3}
  + \frac{\partial x^{\bar{2}}}{\partial x^i}
    D_{q_2 r}^{\bar{2}} u_{q_1 r q_3}             
  + \frac{\partial x^{\bar{3}}}{\partial x^i}
    D_{q_3 r}^{\bar{3}} u_{q_1 q_2 r}
  \Big) && \nonumber\\
= s_{q_1 q_2 q_3}
 -\frac{\sqrt{\bar{\gamma}^{\bar{1}\bar{1}}_{q_1 q_2 q_3}}}{w_{q_1}}
  G_{q_1 q_2 q_3} (\delta_{q_1 0} + \delta_{q_1 N}) && \nonumber\\
 -\frac{\sqrt{\bar{\gamma}^{\bar{2}\bar{2}}_{q_1 q_2 q_3}}}{w_{q_2}}
  G_{q_1 q_2 q_3} (\delta_{q_2 0} + \delta_{q_2 N}) && \nonumber\\
 -\frac{\sqrt{\bar{\gamma}^{\bar{3}\bar{3}}_{q_1 q_2 q_3}}}{w_{q_3}}
  G_{q_1 q_2 q_3} (\delta_{q_3 0} + \delta_{q_3 N}) , \nonumber\\
\ea
where $G$ is defined by
\be
\label{def_G}
G := (n_i A^i u)^* - n_i A^i u .
\ee
When we compare with the surface term $F$ defined in Eq.~(\ref{def_F}),
appearing in the analog Eq.~(\ref{Integ4}), we see that $G$ can be obtained
from $F$ if we replace $f^i$ by $A^i u$.

\subsection{The numerical flux}
\label{numflux}

In the interior of the domain, the flux vector $f^i(u)$ is simply computed
from the field values $u$ in the interior. However, the numerical flux that
is used in the surface integral over the boundary is computed from the field
values on both sides of the boundary. In many cases, we will use the Rusanov
or local Lax-Friedrichs (LLF) flux. It is given by
\be
\label{LLF_flux}
(f^i n_i)^* = \frac{1}{2}\left[
             f^i(u_\mathrm{in}) n_i + f^i(u_\mathrm{adj}) n_i
             + |\lambda|_\mathrm{max}\left(u_\mathrm{in} - u_\mathrm{adj}\right) \right] .
\ee
Here $n_i$ is the outward pointing normal to the boundary, $u_\mathrm{in}$ is the
field value at the boundary using grid points that belong to the domain
enclosed by the boundary, $u_\mathrm{adj}$ is the
field value at the boundary using grid points that belong to the adjacent
domain on the other side of the boundary, and $|\lambda|_\mathrm{max}$ is the
absolute value of the eigenvalue of the characteristic mode
with the largest eigenvalue magnitude, considering eigenvalues from both
sides.

In the case where our system of equations takes the form of
Eq.~(\ref{NonCons}), we often also use another numerical flux, called the
upwind flux. It is constructed from the orthonormalized eigenvectors and
eigenvalues of the matrix $A^i n_i$ appearing in the surface
term~(\ref{def_G}). Let the matrix $S$ contain the eigenvectors as its
columns. Then we can write
\be
A^i n_i = S \Lambda S^{-1} ,
\ee
where $\Lambda$ is a diagonal matrix that contains the corresponding
eigenvalues. The eigenvectors with positive eigenvalues correspond to
modes going along the direction on $n_i$, while the ones with negative
eigenvalues correspond to modes going in the opposite direction. This means
positive and negative eigenvalues are associated with modes that are
outgoing and incoming through the boundary of the domain. As is usually the
case, we wish to impose conditions only on the incoming modes. So, we define
the upwind numerical flux that appears in Eq.~(\ref{def_G}), as
\be
\label{upwind_flux}
(n_i A^i u)^* = (S (\Lambda^+ + \Lambda^-) S^{-1}u)^*
             := S (\Lambda^+S^{-1}u_\mathrm{in} + \Lambda^-S^{-1}u_\mathrm{adj}) .
\ee
Here $\Lambda = \Lambda^+ + \Lambda^-$ with $\Lambda^+$ and $\Lambda^-$
containing the positive and negative eigenvalues, and
$u_\mathrm{in}$ and $u_\mathrm{adj}$ are the
field values from the current domain and the adjacent domain.

\subsection{Patches}

To write Eq.~(\ref{Cons2}) or (\ref{NonCons}), we use particular coordinates
that are chosen to be Cartesian-like, and we call them $x^i = (x,y,z)$ in
\nmesh. As already explained before we map these globally used Cartesian
coordinates to local coordinates $x^{\bar{i}}$ via Eq.~(\ref{x_of_Xbar}).
This mapping is usually carried out in two steps. We first map them into a
particular region or patch via
\be
\label{patch_coords}
x^i = x^i(X^j) .
\ee
For example, we can use standard spherical
coordinates $X^i = (r,\theta,\varphi)$ with a range
$r\in [r_\mathrm{min},r_\mathrm{max}]$,
$\theta\in [\theta_\mathrm{min},\theta_\mathrm{max}]$,
$\varphi\in [\varphi_\mathrm{min},\varphi_\mathrm{max}]$,
so that we cover a certain section of a shell. Next we use
\be
X^i = \frac{1}{2}\left[(X^i_\mathrm{max} - X^i_\mathrm{min}) \bar{X}^i
                  + X^i_\mathrm{max} + X^i_\mathrm{min}\right]
\ee
to map each $X^i$ into an $\bar{X}^i$ that has the standard range
$\bar{X}^i\in[-1,1]$. These $\bar{X}^i$ are what have been denoted by
$x^{\bar{i}}$ in Eq.~(\ref{x_of_Xbar}). Each patch is thus described
by the particular transformation~(\ref{patch_coords}) and range we use
for the $X^i$ coordinates. In some cases, we only need Cartesian coordinates
so that we use the identity transformation in Eq.~(\ref{patch_coords}),
but we have also implemented the transformation to the cubed sphere
coordinates $X^i=(\lambda,A,B)$ described in~\cite{Tichy:2019ouu}.
We then arrange our various patches such that they touch and cover
the region of interest.
\begin{figure}
\includegraphics[width=0.5\linewidth,clip=true]{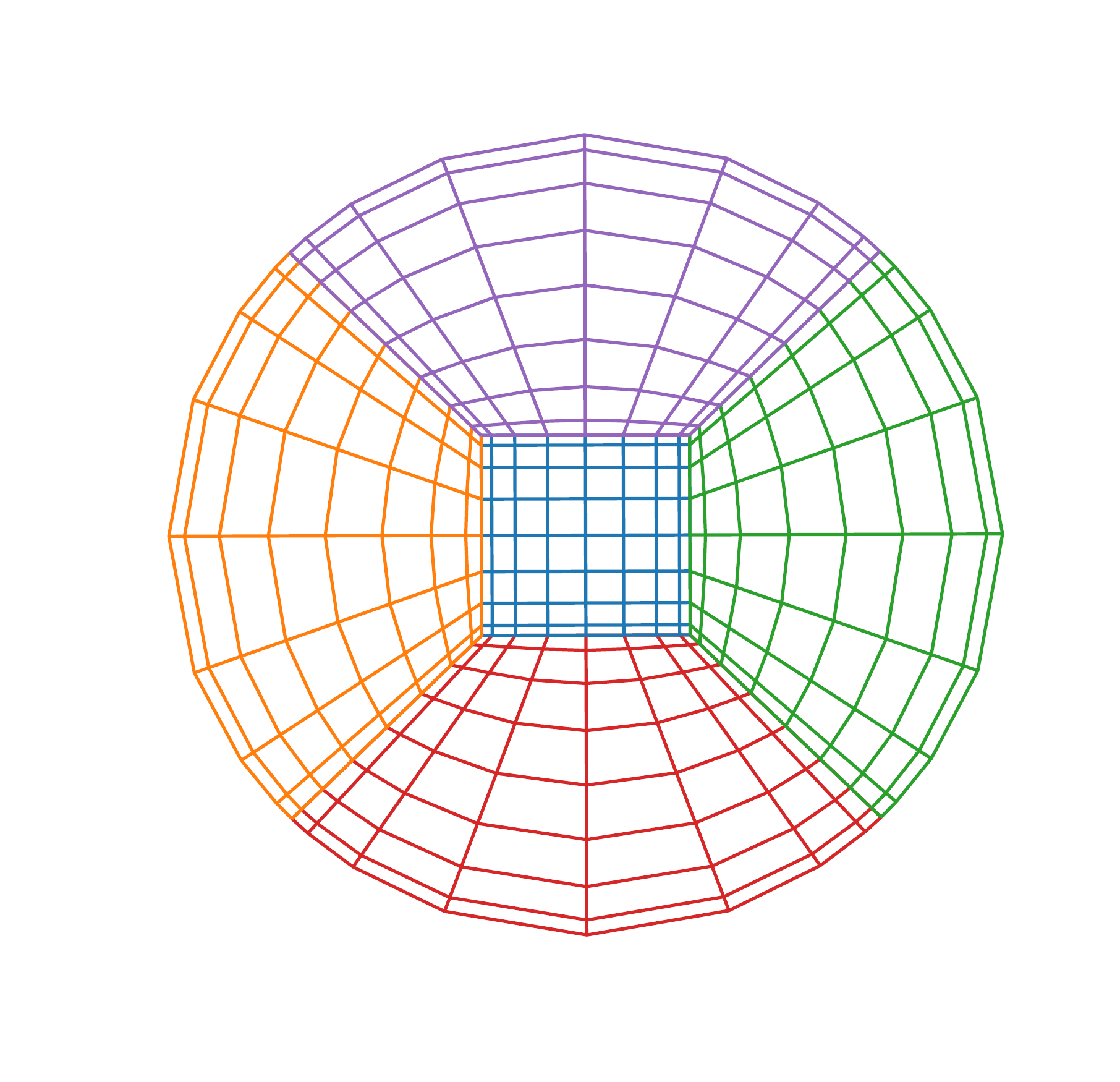}
\includegraphics[width=0.5\linewidth,clip=true]{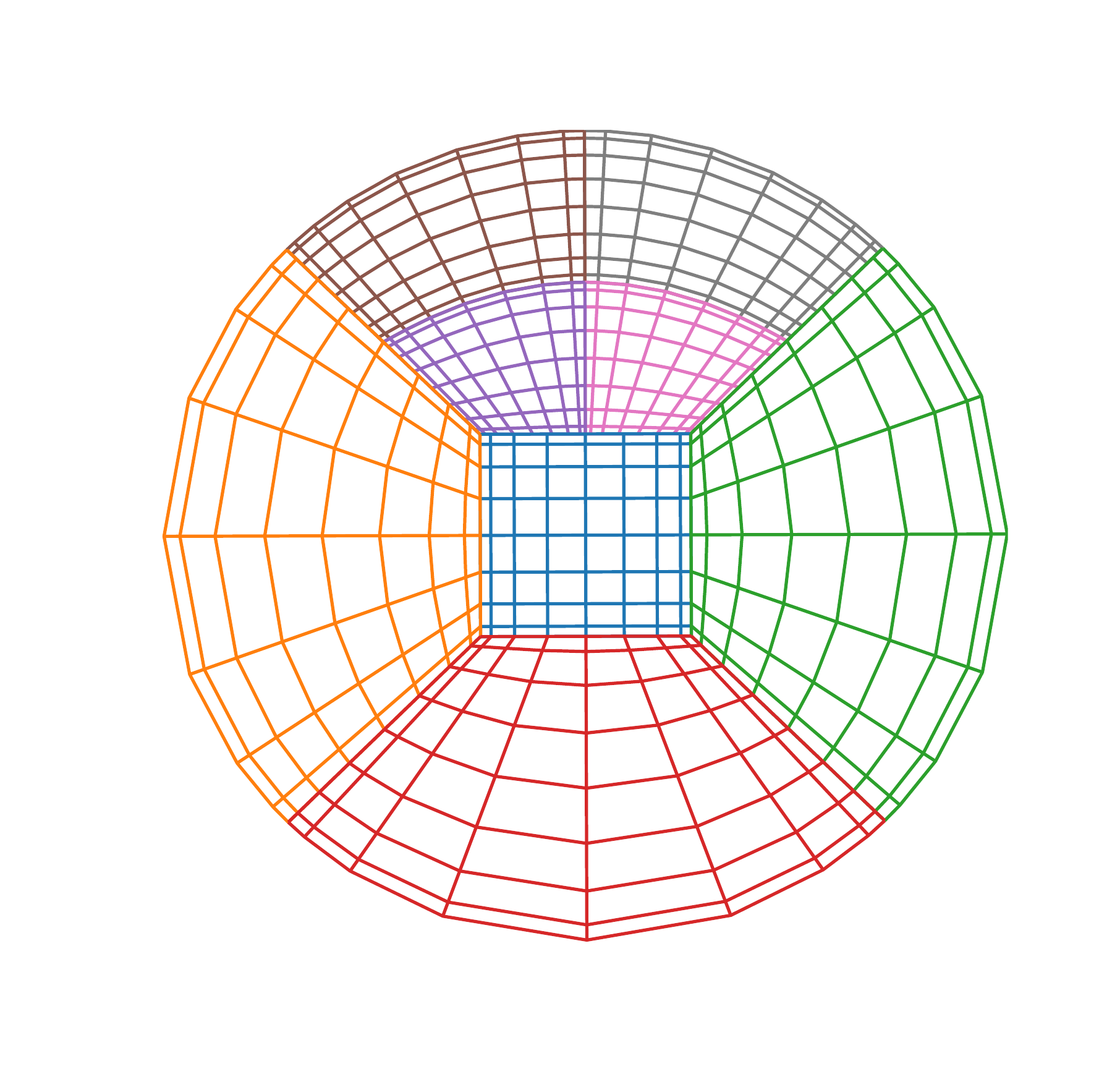}
\caption{\label{CubSphPatches}
On the left side, we show a mesh which is made up of six cubed sphere
patches that are arranged around one central cube. Five of these seven
patches intersect the $xy$-plane and are shown in the picture.
The right side shows the same patches as on the left. However, the
top root node (that covers the entire top patch) has been h-refined so that
this patch is
now covered by eight child nodes, of which we show four in the $xy$-plane.
}
\end{figure}
An example is shown on the left side of Fig.~\ref{CubSphPatches}.
Here we have one central
cube that is covered by Cartesian coordinates. This cube is surrounded
by six cubed sphere patches. Five of these seven patches intersect
the $xy$-plane and are shown on the left of Fig.~\ref{CubSphPatches}.
Note that each of the six cubed sphere patches shares one
face with the central cube. The face on the opposite side is curved and
arises by deforming one side of a larger cube into a spherical surface
via a coordinate transformation of the form in Eq.~(\ref{patch_coords}).
It thus comprises one sixth of the spherical outer boundary.
The remaining faces of each cubed sphere patch touch four other cubed
sphere patches, along flat surfaces. All patches are touching each other
without any overlap, so that all interior patch faces have their entire face
in common with one other patch face. In the next section we explain
how information is exchanged via numerical fluxes between adjacent patches.

\subsection{Adaptive Mesh Refinement}
\label{AMR_sec}

The patches described before can be directly covered with the Legendre
Gau\ss-Lobatto grid points introduced above. Yet, to gain more flexibility, we
can further refine each patch in \nmesh. This is achieved by identifying each
patch with a so-called root node that can be further refined. When we refine
this root node, we cut the original ranges of all three $X^i$-coordinates in
half so that we end up with eight touching child nodes that now cover the
original root node or patch. Each of these new nodes can then be further
refined by again dividing it into eight child nodes. In this way we can
refine each patch as often as we want. This can be done in an irregular way,
where we further refine only the nodes in certain regions of interest. We
end up with a node tree called an octree, where each node has either zero or
eight child nodes. The nodes without children are called leaf nodes.
Together, these leaf nodes cover the entire patch and are thus the nodes in
which we perform any calculations. For this reason, the leaf nodes are often
called computational elements or just elements. However, in this paper, we
will simply call them leaf nodes or just nodes. We note that, in the context
of finite volume methods, the word ``node'' is also sometimes used in the
literature to denote a grid point. Yet, in this paper the word ``node'' will
always refer to a node in our octree.

The $X^i$-range of each leaf node is covered by grid points that correspond
to the Legendre Gau\ss-Lobatto points discussed above. This means that we
have grid points on each node face. This simplifies any calculations
that depend on the values of fields on both sides of a node boundary.
The number of grid points in each node can be freely chosen. When we
increase it, we obtain higher order accuracy if the fields are smooth within
the node. This is called p-refinement because increasing the number of
grid points corresponds to an increase in the number of basis polynomials
we use to represent a field within a node. Of course, p-refinement
is most useful for smooth fields. For non-smooth fields it is often better
to refine a node by splitting it into eight child nodes, which is
known as h-refinement as it refines the resolution even if each child node
has still the same number of grid points as the parent node. In \nmesh~both
p- and h-refinement can be performed whenever desired. Together we call
this Adaptive Mesh Refinement (AMR).


On the right of Fig.~\ref{CubSphPatches}, we show an example where we
h-refine the top node from the left side of Fig.~\ref{CubSphPatches}.
The resulting child nodes now cover the top patch.
As we can see, the grid points of the h-refined nodes along e.g.~the
left patch boundary no longer all coincide with the grid
points of the unrefined node covering the left patch. This is a general
phenomenon, whenever two neighboring nodes differ in their h- or
p-refinement, many of the surface grid points of one node do not coincide
with the surface grid points of the touching adjacent node. Furthermore, the
surface of one node may be touching several adjacent nodes, as is the case
for the node covering the left (orange) patch. This complicates the
calculation of numerical fluxes such as Eq.~(\ref{LLF_flux}) or
Eq.~(\ref{upwind_flux}) in one of our nodes, because we need both the fields
$u_\mathrm{in}$ and $u_\mathrm{adj}$ at every surface grid point of the
current node. We already have $u_\mathrm{in}$ at every point of the current
node. However, $u_\mathrm{adj}$ only exists at the grid points of the
adjacent nodes, which in general do not coincide with the surface grid
points of the current node. To obtain $u_\mathrm{adj}$ at one of the
surface grid points of the current node, we interpolate the $u_\mathrm{adj}$
data from the adjacent node onto this point. For this we currently use
Lagrange interpolating polynomials constructed from the 2-dimensional
surface data of the adjacent node. To easily find adjacent neighbors each
node has an associated data structure that keeps track of all adjacent
neighbor nodes. When p-refinement is applied to a node this data structure
can remain unchanged since the size of the nodes and therefore the number
of neighbors does not change. However, the data structure has to be updated
whenever a node is h-refined. In this case the structure gets updated on
this one node and on all of its neighbors. In this way \nmesh~is able to
accommodate arbitrary levels of h-refinement. For example, it is possible to
h-refine a node into eight child nodes, and to then repeat this as often as
desired with any of the child nodes, without at the same time refining any of
the original neighbor nodes.
In each case the final result is a number of touching leaf nodes that cover
each patch. Since the patches themselves are also touching, the collection
of all leaf nodes from all patches forms the mesh on which we perform our
calculations.

The word Adaptive in AMR usually also implies an algorithm that
automatically chooses p- or h-refinement. We currently have implemented only
one such algorithm. Within each node, it expands a chosen quantity, such as
the matter density $\rho_0$, in terms of Legendre polynomials. The
coefficients in front of each of the Legendre polynomials can then be used
to judge the smoothness of $\rho_0$. If $\rho_0$ is perfectly smooth, we
expect the coefficients to fall off exponentially with increasing polynomial
order. In our algorithm we then fit the logarithm of the coefficient
magnitudes to a linear function. If the slope values $b_i$ in all three
directions ($i=1,2,3$) of this linear function are not negative enough,
we consider $\rho_0$ to be not smooth. We then h-refine the node. This
algorithm is in principle geared toward dealing with the non-smooth behavior
of matter across a neutron star surface. We have, however, not had any real
success with this algorithm yet. We can turn it on and evolve neutron stars
with it, but we have not been able to tune the parameters, that decide
when the $b_i$ are not negative enough, to values that work well after some
matter leaves the star surface. We mention this particular algorithm here
only to show that \nmesh~has AMR capabilities in principle. We note,
however, that these capabilities were not used in the simulations discussed
below, where we use uniform h-refinement.
\begin{figure}
\includegraphics[width=1\linewidth,clip=true]{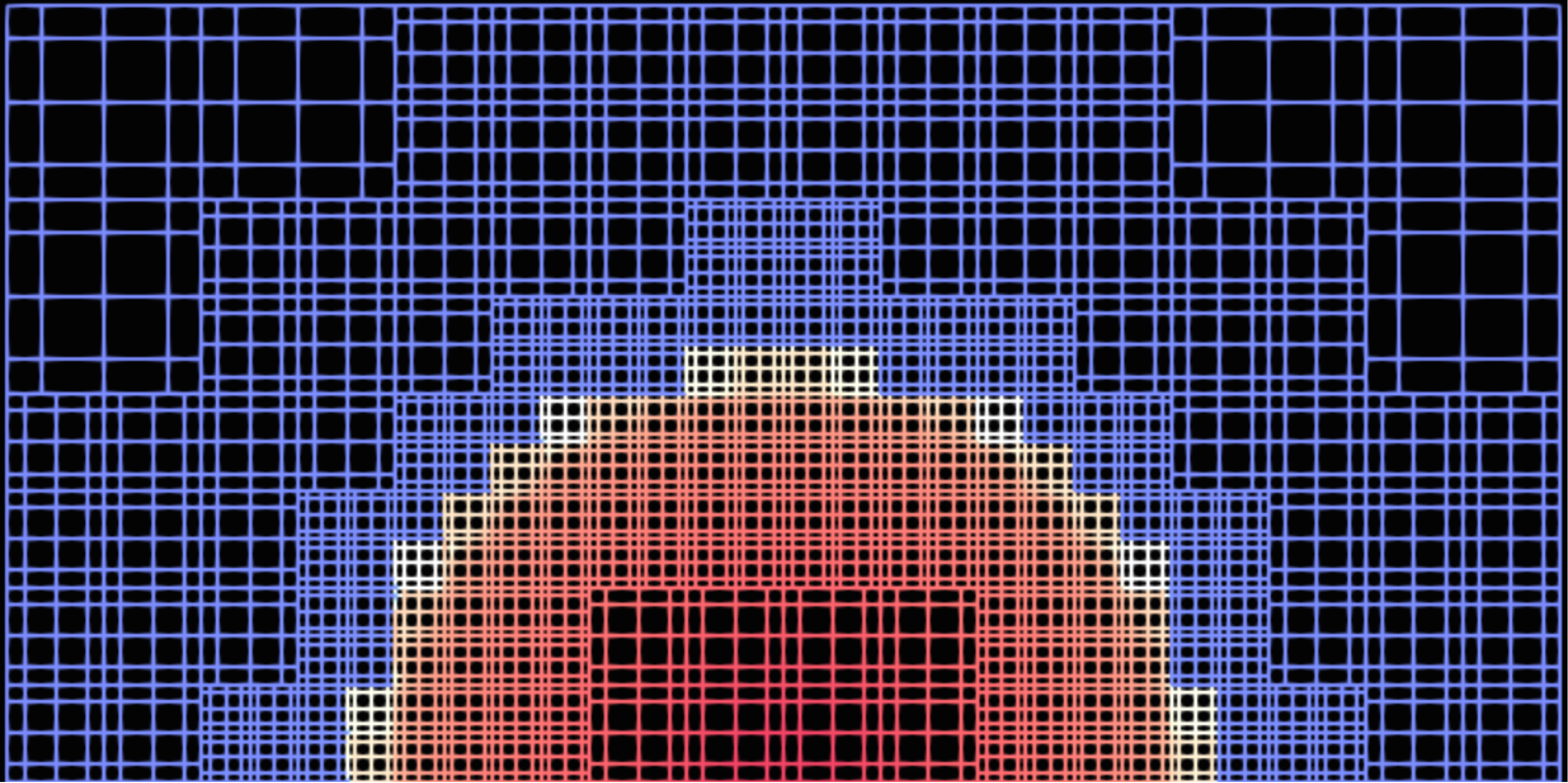}
\caption{\label{TOVref}
Leaf nodes near the neutron star surface are most refined. Inside the
star (red region) and outside the star (blue region) less refinement is
used.
}
\end{figure}
Figure.~\ref{TOVref} shows the mesh for a single neutron star in a plane
through its center. The different levels of h-refinement shown here are
obtained from the coefficient drop off based algorithm mentioned above.
It also refines neighbor nodes if their level of refinement is
more than one below the just refined node. The purpose of the latter is to
avoid abrupt changes in resolution, but is technically not required by
\nmesh.

\subsection{Time integration}

Note that Eqs.~(\ref{Integ4}) or (\ref{IntegA}) still contain time
derivatives, since up to this point we have only discretized spatial
derivatives. This means that these equations represent a set of coupled
ordinary differential equations for the fields $u_{q_1 q_2 q_3}$ at
the grid points. In this paper, we use standard Runge-Kutta time integrators
to find the solution of these ODEs from the $u_{q_1 q_2 q_3}$ at the initial
time. Such Runge-Kutta methods are only stable when the
Courant-Friedrichs-Lewy (CFL) condition is satisfied, i.e., if the time step
$\Delta t$ is small enough. In this work we use
\be
\label{cfl}
\Delta t = \Delta x_\mathrm{min} / v ,
\ee
where $\Delta x_\mathrm{min}$ is the distance between the two closest grid points,
and $v$ is a number that needs to be greater than the largest characteristic
speed.

\subsection{Filters that can improve stability}

Even when the time step satisfies the CFL condition, instabilities can still
occur in some cases, e.g., on non-Cartesian patches. To combat such
instabilities, we filter out high frequency modes in the evolved fields.
This is achieved by first computing the coefficients $c_{l_1 l_2 l_3}$ in
the expansion
\be
u(\bar{x})J
= \sum_{l_1=0}^N\sum_{l_2=0}^N\sum_{l_3=0}^N c_{l_1 l_2 l_3}
   P_{l_1}(x^{\bar{1}}) P_{l_2}(x^{\bar{2}}) P_{l_3}(x^{\bar{3}})
\ee
of each field $u$, where the $P_{l}(\bar{x})$ are Legendre polynomials.
The coefficients are then replaced by
\be
\label{exp_filter}
c_{l_1 l_2 l_3} \to
c_{l_1 l_2 l_3} e^{-\alpha_\mathrm{f} (l_1/N)^s} e^{-\alpha_\mathrm{f} (l_2/N)^s}
                e^{-\alpha_\mathrm{f} (l_3/N)^s} ,
\ee
and $u$ is recomputed using these new coefficients. Note that we
typically use $\alpha_\mathrm{f}=36$ and $s=32$, so that the coefficients with the
highest $l=N$ are practically set to zero, while all others are mostly
unchanged.

\section{Parallelization strategy}
\label{parallelization_strategy}

Modern supercomputers are made of thousands of compute nodes, each with on
the order of 100 Central Processing Unit (CPU) cores. Each compute node has
its own separate memory (typically on the order of 100 GB), and cannot
directly access data on other compute nodes. However, all compute nodes are
connected by a network that allows data transfers between them. The by now
traditional way to parallelize programs on such supercomputers is to use the
Message Passing Interface (MPI) library. With MPI, we start multiple
processes (i.e., programs), each using its own piece of memory. Typically
each process then works on a part of the problem that we wish to solve. The
only way to exchange data is via messages sent between the different
processes, hence the name MPI. Since no direct memory access occurs, MPI
works very well if the processes run on different compute nodes that do not
share any memory. Nevertheless, it is also possible to start multiple MPI
processes within one compute node to take advantage of the presence of
multiple CPU cores.

Systems consisting of black holes or neutron stars are governed by partial
differential equations. To discretize them, we use the DG method together
with AMR, as described above, so that the region of interest is covered by
a number of leaf nodes (as described in Sec.~\ref{AMR_sec}).
We typically use several levels of
h-refinement so that we end up with a large number of leaf nodes, possibly
hundreds of thousands or even millions. The parallelization strategy of
\nmesh~is then to distribute these leaf nodes (referred to simply as nodes
below) among the available MPI processes.
To take one time step, we need to evaluate the various terms in
Eq.~(\ref{Integ4}) or (\ref{IntegA}). Notice that $F$ and $G$ in these
equations depend on field values from the surface points of adjacent nodes
via the numerical flux. Hence, MPI messages need to be sent to obtain these
surface values. All other terms in Eqs.~(\ref{Integ4}) and (\ref{IntegA})
depend only on field values local to each node. Thus, we instruct MPI to
start the transfer of the surface values. While this transfer is ongoing, we
locally calculate all the terms in Eq.~(\ref{Integ4}) or (\ref{IntegA})
besides $F$ or $G$. This allows us to overlap communication and calculation,
i.e., we avoid waiting for network transfers to arrive. Also note that the
amount of data that needs to be sent via MPI is quite small, as the only
values that need to be exchanged are from points on the surfaces of adjacent
nodes. This is a significant advantage of DG methods compared to more
traditional finite difference or finite volume methods. The latter two
require transfer of data from a layer several points deep. The depth of this
layer even increases when one increases the order of accuracy of the finite
difference or finite volume method. Furthermore, if one uses coordinate
patches, such as the cubed spheres as discussed above, one even needs data
from more than just the six directly adjacent neighbor nodes (cf.
\cite{Bugner2018a}), because if we go several points deep in a curved
coordinate direction, we may end up in yet another node. We thus expect our
DG method to be more efficient.

\begin{figure}
\centering
\includegraphics[width=1\linewidth]{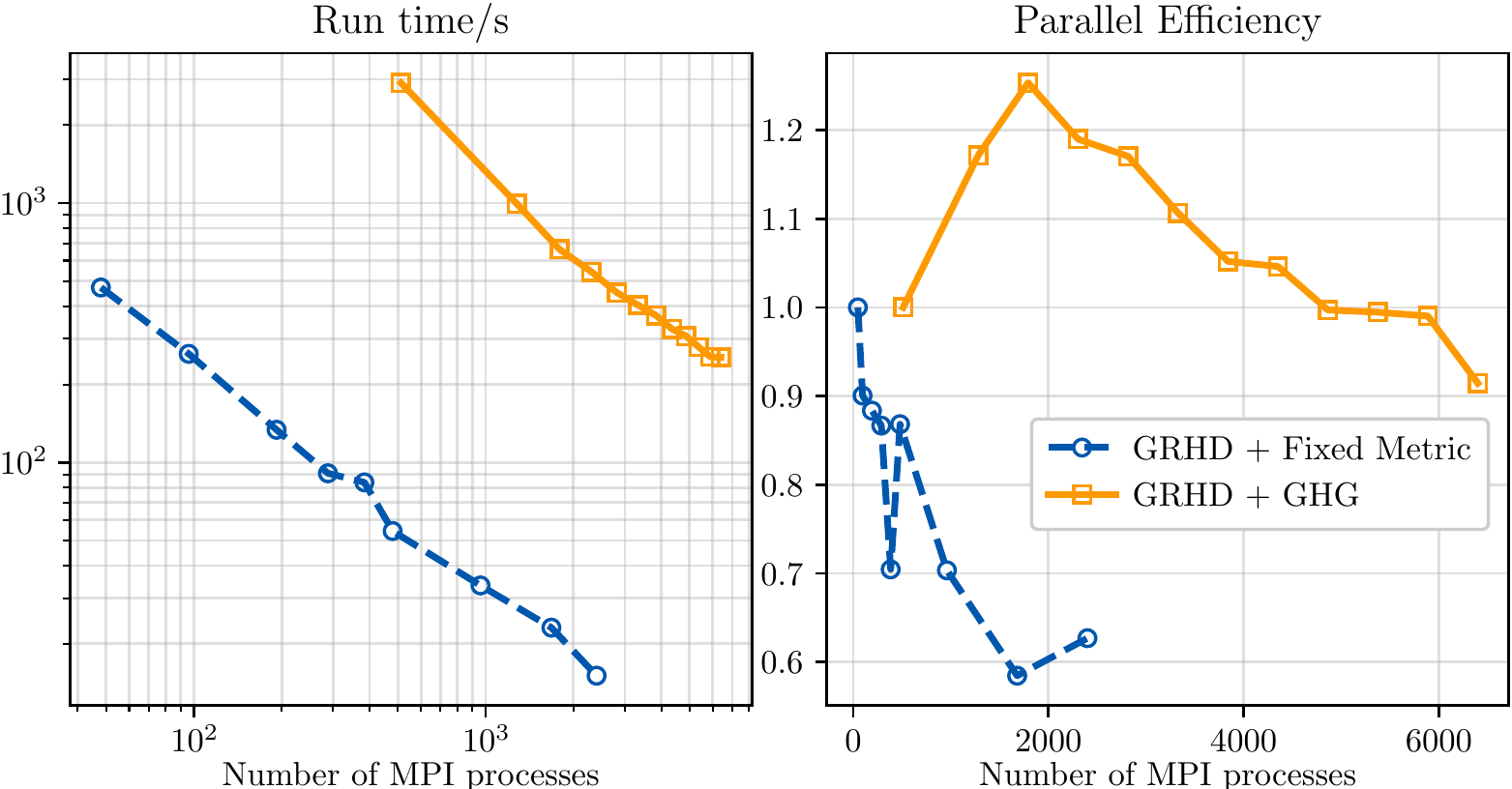}
\caption{
\label{strongscaling}
\small{
Strong scaling tests for the evolution of a neutron star using 262144 leaf
nodes with $5\times5\times5$ points. Circles indicate results obtained on
the Cartesius supercomputer for a fixed spacetime metric. Squares show
results on the Bridges-2 supercomputer, where the metric is evolved as well.
}
}
\end{figure}
To demonstrate the efficiency of \nmesh~we have performed two strong scaling
tests. On the left side of Fig.~\ref{strongscaling} we show the run time vs
the number of MPI processes used. The circles correspond to the simulation
of a single neutron star on a fixed spacetime metric on the Cartesius
supercomputer. The squares are from the simulation of a single neutron star
together with an evolving spacetime metric, that was performed on the
Bridges-2 supercomputer. Perfect scaling corresponds to a linear speedup as
this fixed size problem is evolved with more MPI processes.
As we see on the left side of Fig.~\ref{strongscaling}, our run time
measurements almost follow a straight line, and thus indicate good scaling.
To study the scaling further we also show the parallel efficiency on the
right side of Fig.~\ref{strongscaling}. This efficiency is computed from
$t_r(1)/[n t_r(n)]$, where $t_r(n)$ is the run time measured using $n$ MPI
processes. Since a single MPI process cannot obtain enough memory for the
simulations, $t_r(1)$ is estimated from
$t_r(1) = n_{\mathrm{min}} t_r(n_{\mathrm{min}})$, where $n_{\mathrm{min}}$
is the run with the lowest number of MPI processes performed in each case.

Perfect scaling would correspond to a constant parallel efficiency. However,
this is typically not achieved by real programs. In the case of \nmesh~any
sort of scaling will definitely stop once the number of MPI processes
becomes comparable to the number of leaf nodes (here 262144). In fact, we
expect it to stop even before this, due to the growing communication
overhead when more parallelization is used. The fact that the run with the
evolving spacetime metric has a higher efficiency might be related to the
fact that the evolution of the metric is time consuming, so that every MPI
process does more work before communication is needed again. On the other
hand, Bridges-2 had newer hardware and MPI libraries than Cartesius, which
could also account for part of the difference. The fact that
our curves end at 2400 and 6400 MPI processes, is not due to any
particular limitation of \nmesh. Rather, we currently do not have access to
a machine that would allow us to use more cores. It thus remains to be seen
up to which number of cores \nmesh~will scale well. Nevertheless we consider
our results encouraging. They confirm the expectation that a program based
on a DG method should have good scaling.

\section{Evolution system tests and results}
\label{evo_system_tests}

In this section, we perform tests with several different evolution systems to
validate our new \nmesh~program. We also explain in detail which methods we
use for our simulations of general relativistic hydrodynamics, and then
show our results.

%
%

\subsection{Scalar wave equation}

One of the simplest systems one can evolve is a scalar wave. Here we
consider a single scalar field obeying the wave equation
\be
\partial_t^2\phi = \delta^{ij} \partial_i\partial_j\phi .
\ee
The DG method described earlier cannot be applied directly to systems
with second order derivatives. We therefore introduce the extra variables
\be
\Pi := \partial_t\phi
\ee
and
\be
\chi_i := \partial_i\phi .
\ee
This results in the following system of first order equations
\ba
\label{scalar_evo}
\partial_t\Pi    + \partial_j f_{\Pi}^j    &=&  0,   \nonumber \\
\partial_t\chi_i + \partial_j f_{\chi_i}^j &=&  0,   \nonumber \\
\partial_t\phi                             &=&  \Pi.
\ea
Here we have defined the flux vectors
\ba
f_{\Pi}^j    &:=& -\chi_j,        \nonumber \\
f_{\chi_i}^j &:=& -\Pi\delta_i^j, \nonumber \\
f_{\phi}^j   &:=& 0.
\ea
As we can see, the system in Eq.~(\ref{scalar_evo}), consists of two coupled
partial differential equations and one ordinary differential equation.

To evolve this system with our DG method, we also need to provide initial
values and boundary conditions. Since
\be
\label{analytic_phi}
\phi = \sin(k_i x^i - \omega t)
\ee
is a solution for any $k_i$ and $\omega = \sqrt{\delta^{ij}k_i k_j}$, we
initialize the system according to this equation at $t=0$. We also use
Eq.~(\ref{analytic_phi}) at the outer boundary so that this sine wave is
continuously entering through the outer boundary.

The DG method requires numerical fluxes. We have successfully used both the
LLF flux of Eq.~(\ref{LLF_flux}) as well as the upwind flux
of Eq.~(\ref{upwind_flux}).
To impose Eq.~(\ref{analytic_phi}) at the outer boundary, we use the same
numerical flux as in the interior, but we set $u_\mathrm{adj}$ to the value coming
from Eq.~(\ref{analytic_phi}).

\begin{figure}
\includegraphics[width=\linewidth]{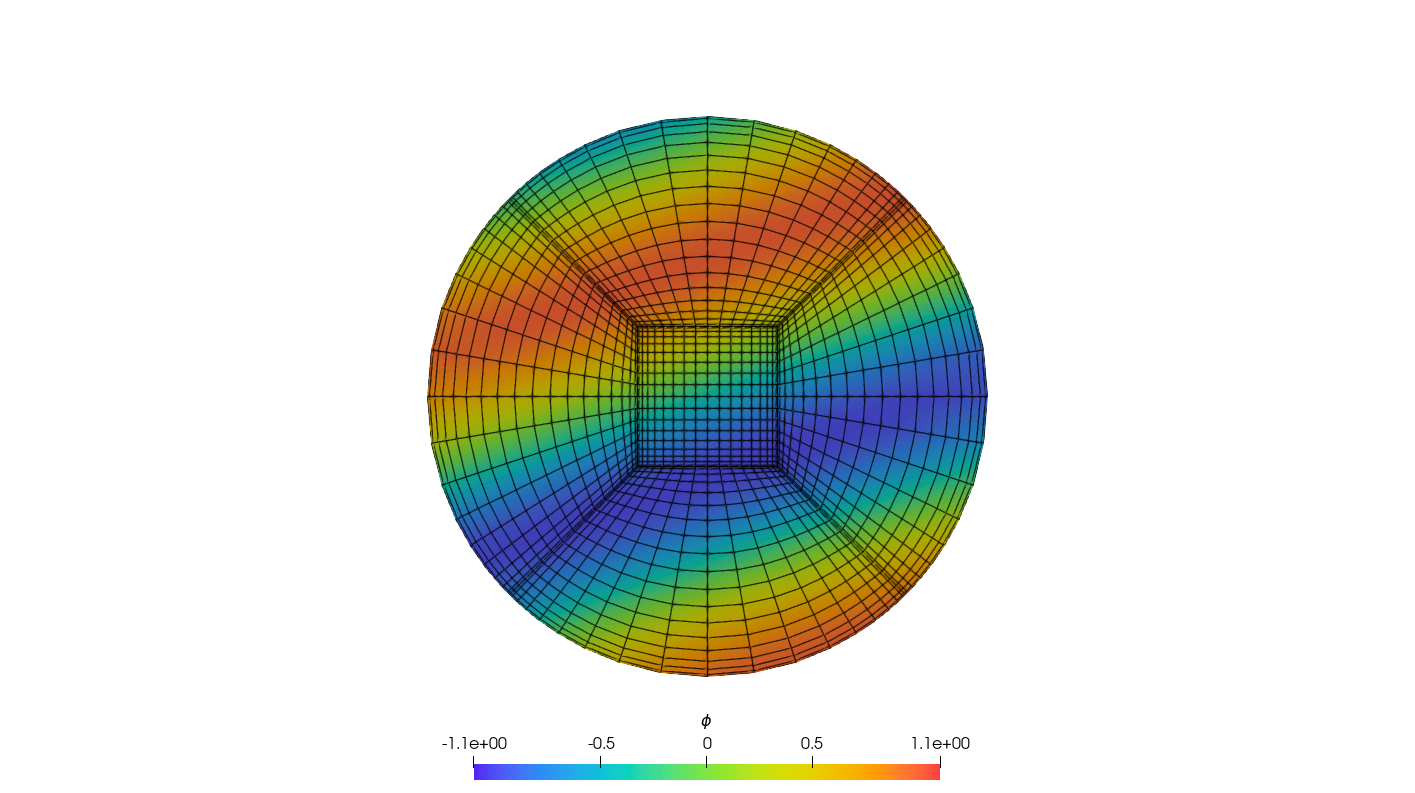}
\caption{\label{scwav-grid}
The plot shows the scalar field $\phi$ in terms of a colormap together with
the mesh (in black) at evolution time $t = 4.5$.
The mesh is made up of seven leaf nodes, five of which intersect the
$xy$-plane shown here.
}
\end{figure}
The wave vector $k_i$ in Eq.~(\ref{analytic_phi}) is arbitrary.
For the test results presented here, we choose $k_i=(0.7, -2, 4.3)$,
so that it represents the general case where $k_i$ is not aligned
with any coordinate direction.
As we can see in Fig.~\ref{scwav-grid}, we get a sinusoidal wave that
propagates through our numerical domain, with this $k_i$ vector.
In this case we have used seven patches and the figure
shows the scalar field $\phi$ in the $xy$-plane after evolving up to
time $t = 4.5$. For the test case depicted in
Fig.~\ref{scwav-grid}, we have used an equal number of grid
points ($19 \times 19 \times 19$) in all directions,
without any h-refinement applied to the root nodes.
However, we have also performed tests with an unequal number
of grid points and with the root nodes h-refined.
We have obtained stable evolution for the system for all these
cases with both the LLF and upwind numerical fluxes. The choice of
numerical flux did not have any significant effect in any of the
scalar wave evolution tests, as both cases yield results that have
errors of the same order.

\begin{figure}
\includegraphics{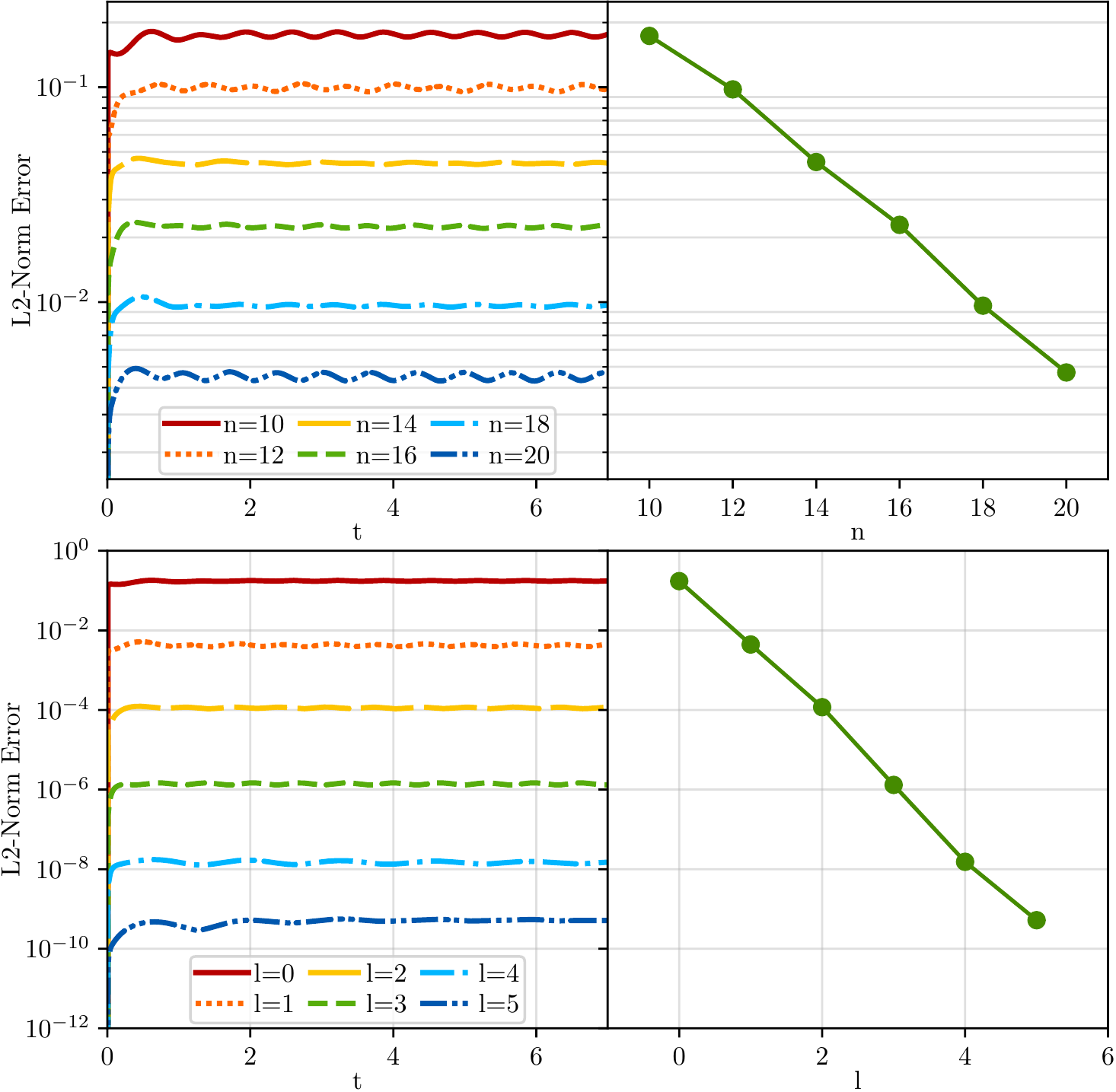}
\caption{\label{scwav-conv}
This plot shows the change in the L2-norm of the error in the scalar field
$\phi$ for two different cases when using the LLF numerical flux.
First, p-refinement is applied (without any h-refinement)
and we plot the error vs time (top left) and then the error vs number of
points $n$ along each axis in a node at time $t = 3$ (top right).
Next, h-refinement is applied, and we plot the error vs time (bottom left)
and then the error vs levels of refinement $l$ applied to the root node
at time $t = 3$ (bottom right). In this case, we keep the number of
points in each direction in each node fixed at $n=10$.
}
\end{figure}
To demonstrate the convergence of our new \nmesh~program in this
scalar wave evolution test, we have applied p-refinement and h-refinement
separately. In Fig.~\ref{scwav-conv} we plot the L2-norm of the error
in $\phi$ for both. For the case of p-refinement (top)
we increase the number of points $n$ in all directions,
setting them to $n = 10,12,14,16,18,20$, with no h-refinement
applied to the root nodes. We observe an
exponential drop in the L2-norm error, as expected in this case.
For the case of h-refinement (bottom), we apply $l = 0,1,2,3,4,5$
levels of refinement to the root nodes, and have $n=10$
points in each direction in each node. Again, we observe
convergent behavior for the L2-norm of the error, when increasing
the number of h-refinement levels. Figure~\ref{scwav-conv} only shows
the results for the LLF numerical flux, since the results
for the upwind flux are very similar. For all the runs shown in
Figs.~\ref{scwav-grid} and \ref{scwav-conv}, the time step was set to
$\Delta t = \Delta x_\mathrm{min} / 5$ in accordance with Eq.~\ref{cfl}.

Even though all the convergence results stated above have been obtained
for a mesh using six cubed sphere patches that surround one central cube,
we also obtain convergence for a mesh covered by a single cubic Cartesian
patch. The key difference between these is that we need
the filters of Eq.~\ref{exp_filter} to
stabilize the evolution on meshes that contain both Cartesian and
cubed sphere patches, while this is not necessary on a purely Cartesian
patch, even if it is h-refined.
For the runs of Figs.~\ref{scwav-grid} and \ref{scwav-conv}
we set the filter parameters of Eq.~\ref{exp_filter} to the
values $\alpha_\mathrm{f}=36$ and $s=32$.

\subsection{Convergence tests with the GHG system for a single excised
black hole}
\label{GHG_runs}

For the gravitational part, we have implemented the first-order reduction of
Generalized Harmonic Gauge (GHG)
formulation~\cite{Lindblom:2005qh,Hilditch:2015aba}
\ba
\label{eq:GHG_g}
&\partial_t g_{ab}&
-(1+\gamma_1)\beta^k\partial_kg_{ab} =
-\alpha\Pi_{ab}-\gamma_1\beta^k\Phi_{kab}, \\
\label{eq:GHG_Pi}
&\partial_t\Pi_{ab}&
-\beta^k\partial_k\Pi_{ab}
+\alpha\gamma^{ki}\partial_k\Phi_{iab}
-\gamma_1\gamma_2\beta^k\partial_kg_{ab} = \nonumber \\
&&2\alpha g^{cd}
\left(\gamma^{ij}\Phi_{ica}\Phi_{jdb}-\Pi_{ca}\Pi_{db}
-g^{ef}\Gamma_{ace}\Gamma_{bdf}\right) \nonumber\\
&&-2\alpha\nabla_{(a}H_{b)}-\frac{1}{2}\alpha n^cn^d\Pi_{cd}\Pi_{ab}
-\alpha n^c\Pi_{ci}\gamma^{ij}\Phi_{jab} \nonumber\\
&&+\alpha\gamma_0\left[2\delta^c{}_{(a}n_{b)}-g_{ab}n^c\right]
\left(H_c+\Gamma_c\right)-\gamma_1\gamma_2\beta^k\Phi_{kab} \nonumber\\
&&-16\pi\alpha
\left(
  T_{ab}-\frac{1}{2}g_{ab}g^{cd}T_{cd}
\right), \\
\label{eq:GHG_Phi}
&\partial_t\Phi_{iab}&
-\beta^k\partial_k\Phi_{iab}+\alpha\partial_i\Pi_{ab}
-\alpha\gamma_2\partial_ig_{ab} = \nonumber \\
&&\frac{1}{2}\alpha n^cn^d\Phi_{icd}\Pi_{ab}+\alpha\gamma^{jk}n^c
\Phi_{ijc}\Phi_{kab}-\alpha\gamma_2\Phi_{iab}.
\ea
Here $g_{ab}$ is the spacetime metric, $n_a$ is the unit normal to the
hypersurface of constant coordinate time $t$, and
$\Gamma_a=g^{bc}\Gamma_{abc}$ is the contracted Christoffel symbol.
The equations are written in terms of the extra variables
$\Pi_{ab}:=-n^c\p_cg_{ab}$ and $\Phi_{iab}:=\p_ig_{ab}$,
that have been introduced to make the original second-order GHG system
first-order in both time and space. Gauge conditions in the GHG system are
specified by prescribing the gauge source function $H_a$.
The lapse $\alpha$, shift $\beta^i$ and spatial metric $\gamma_{ij}$ come
from the $3+1$ decomposition in Eq.~(\ref{4metric}).
The GHG evolution equations also contain extra terms that are multiplied with
the parameters $\gamma_{0}$, $\gamma_{1}$, and $\gamma_{2}$. In this paper
we set $\gamma_{1}=-1$, and choose $\gamma_{2}=\gamma_{0}=1$ for the
constraint damping parameters.

To test the gravitational part of \nmesh, we evolve a black hole spacetime.
As initial data, we choose the metric of a single Schwarzschild black hole
in Kerr-Schild coordinates~\cite{Baumgarte_Shapiro_book},
\be
g_{ab}=\eta_{ab}+\frac{2M}{r}l_al_b, \label{eq:Kerr-Schild}
\ee
where $\eta_{ab}$ is the Minkowski metric, and $M$ is the mass of the black hole.
In the Cartesian coordinates, $r=(x^2+y^2+z^2)^{1/2}$, and $l_a=(1, x/r, y/r, z/r)$.
The gauge source function is initialized based on the above
metric~(\ref{eq:Kerr-Schild}), and is left constant during the
simulation~\cite{Bruegmann:2011zj},
\be
H_a(t=0) = -\Gamma_a(t=0), \quad \partial_tH_a=0.
\ee
With this initial condition, the analytic solution of the evolution equations
is simply the static Schwarzschild metric, so that all evolved fields should
be constant. Of course evolution will lead to some amount of
numerical errors. Thus we test here if \nmesh~can stably evolve this setup and
whether the numerical evolution will settle down to a stable state.

As computational domain, we choose a spherical shell that extends from
$r=1.8M$ to $11.8M$, and is covered by six cubed sphere patches.
The inner boundary is thus inside the black hole horizon of $r=2M$.
The speeds of all characteristic modes at the inner boundary are such that
every mode is moving towards the black hole center and thus leaving the
computational domain. We therefore do not impose any boundary
conditions at the inner boundary. The situation at the outer boundary is
different as we have both incoming and outgoing modes. We impose no condition
on the outgoing modes, but we keep the incoming modes constant at their
initial values (consistent with the static analytic solution).
This is done using Eq.~(\ref{upwind_flux}), where $u_\mathrm{adj}$ is set to the
analytic Schwarzschild solution, $u_\mathrm{in}$ are the evolved fields at the
boundary, and $S$, $\Lambda^+$, and $\Lambda^-$ come from the characteristic
modes and their eigenvalues~\cite{Lindblom:2005qh}, calculated from normals
that are normalized with respect to the flat metric.
To improve the accuracy we use either 2 or 3 levels of h-refinement in each
patch. We choose the time step according to Eq.~(\ref{cfl}), with $v=4$. We
find that, with this setup, no filters are necessary to stabilize our runs.
As in the scalar field test cases discussed above, filters only become
necessary when both Cartesian and cubed sphere patches are present. In the
latter case, the filter of Eq.~(\ref{exp_filter}) is again sufficient for
obtaining stable runs.
We have evolved this setup using both the LLF and upwind fluxes
of Eqs.~\ref{LLF_flux} and \ref{upwind_flux} at inter domain boundaries.
As described below, both fluxes work about equally well.

\begin{figure}
  \centering
  \includegraphics[width=0.95\textwidth]{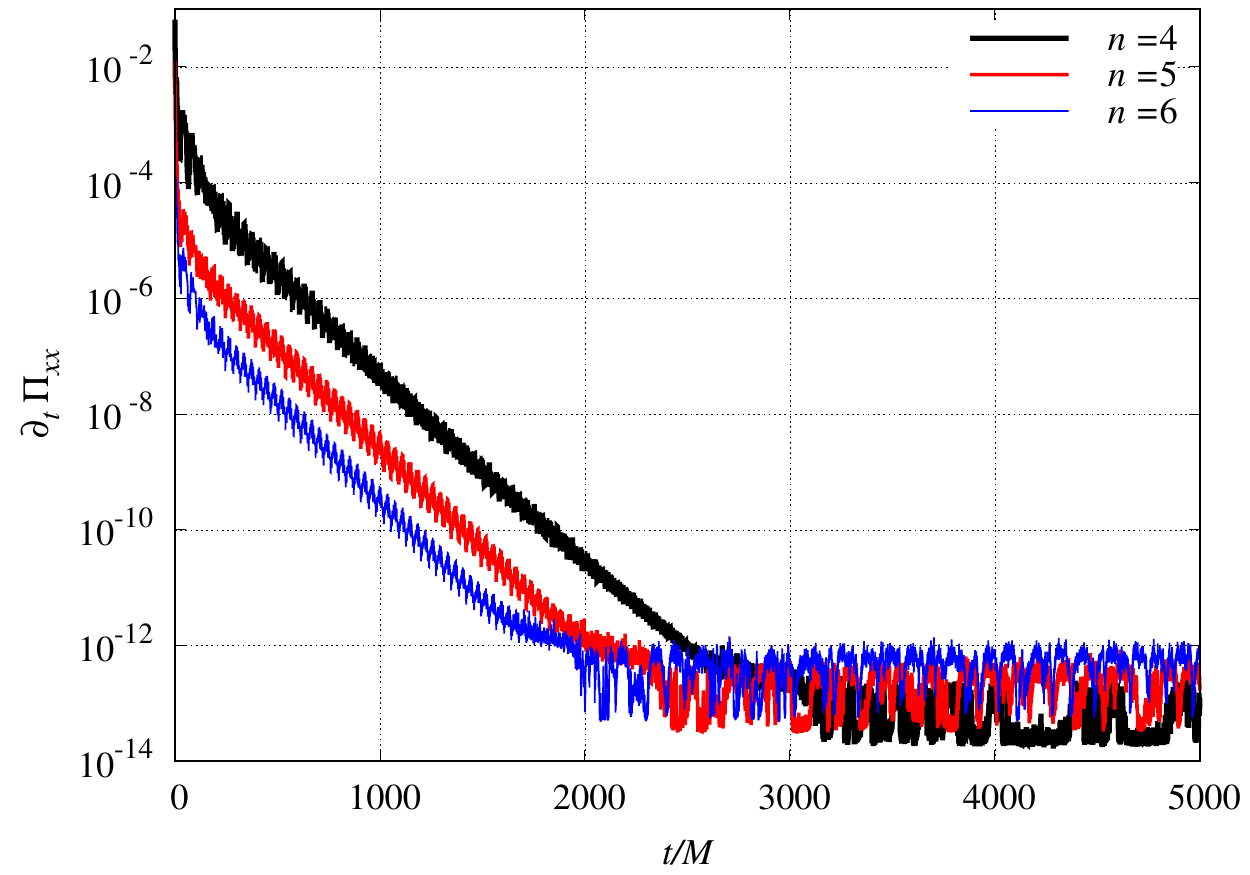}\par
  \caption{\label{KS_stability}
  Time evolution of the time derivative of $\Pi_{xx}$ for a black hole in
  Kerr-Schild coordinates.
  Here, we use $n$ grid points in each direction in each leaf node, and have
  h-refined each of the root nodes three times. We use the upwind flux.
  }
\end{figure}
In order to demonstrate stability of our runs at high resolution, we have
evolved the black hole with three levels of h-refinement for three different
numbers of grid points. We find that the system of equations reaches a state
where the time derivatives $\partial_t g_{ab}$, $\partial_t\Pi_{ab}$, and
$\partial_t\Phi_{iab}$ all approach zero (up to machine precision), as
expected for a static black hole. As an example, we show $\partial_t\Pi_{xx}$
in Fig.~\ref{KS_stability} when evolved with $4\times 4\times 4$, $5\times
5\times 5$, and $6\times 6\times 6$ grid points in each node. As we can see,
this time derivative falls exponentially until it settles down to below
$10^{-12}$. Beyond this point, the terms that determine the time derivatives
in the GHG evolution equations
(\ref{eq:GHG_g}), (\ref{eq:GHG_Pi}), (\ref{eq:GHG_Phi})
add up to almost zero, and deviate from zero only
because of roundoff errors due to the use of floating point numbers.
As expected, at higher resolution this steady state is reached earlier,
because our numerical method then has smaller discretization errors.

\begin{figure}
  \begin{multicols}{2}
    \includegraphics[width=\textwidth]{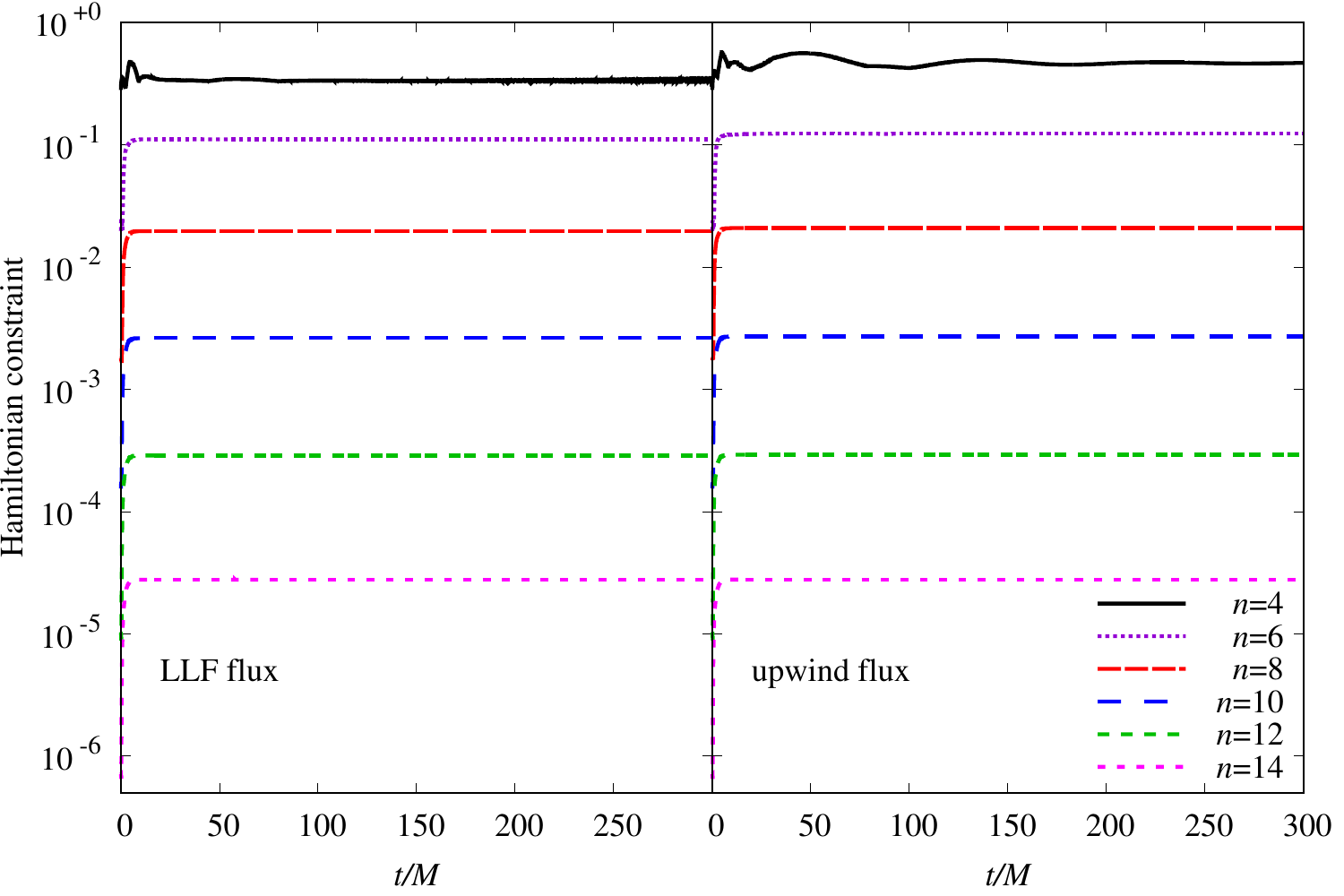}\par
  \end{multicols}
\caption{\label{KS_conv}
  The infinity norm of the Hamiltonian constraint, when evolving a
  black hole in Kerr-Schild coordinates.
  On the left, the constraint evolution is shown, when using the LLF flux.
  On the right are the corresponding results for the upwind flux.
  All the simulations on both sides use grids with $6$ patches, and each
  patch is refined uniformly twice. The different lines correspond to
  different numbers of grid points $n$ in each direction in each leaf node.
  The Hamiltonian constraint converges to zero exponentially as we
  increase $n$.
}
\end{figure}
The Hamiltonian and momentum constraints of general relativity read as
\ba
\label{ham0}
H    &=& R -  K_{ij}  K^{ij} + K^2  - 16\pi\rho_\mathrm{ADM}, \\
\label{mom0}
M^i &=&  D_j(K^{ij} - \gamma^{ij} K) - 8\pi j^i,
\ea
where $R$ and $D_j$ are the Ricci scalar and derivative operator associated
with the 3-metric $\gamma_{ij}$,
$K_{ij} = -\frac{1}{2}\pounds_{n}\gamma_{ij}$,
$\rho_\mathrm{ADM} = T_{ab}n^a n^b$,
and $j^i = -T_{ab}n^a \gamma^{bi}$ (see e.g.~\cite{Tichy:2016vmv}).
General relativity dictates $H = M^i = 0$ for all time. In
Fig.~\ref{KS_conv}, we show the infinity norm of $H$ over the grid when we
evolve the black hole with 2 levels of h-refinement for various numbers of
grid points per node. As we can see, $H$ stabilizes after a short time and
then stays practically constant, which again indicates stability.
As expected $H$ converges to zero exponentially as we
increase the number of grid points. We also see that the results for the
LLF flux (on the left) and the upwind flux (on the right) are almost the
same. In our simulations the momentum constraint $M^i$ behaves just like
$H$ and also converges to zero exponentially as we increase the number
of grid points.

Since the initial data is given by the Kerr-Schild metric, the analytic
solution is a static black hole. The numerical solution, however, evolves
for a while until it settles down into a stable configuration (see
Fig.~\ref{KS_stability}). This happens because the analytic solution does
not exactly satisfy the discretized GHG equations. As already mentioned in
Sec.~\ref{num_methods} we use domain normals that are normalized with
respect to the flat metric for our black hole evolutions.
\begin{figure}
\centering
\includegraphics[width=0.6\linewidth]{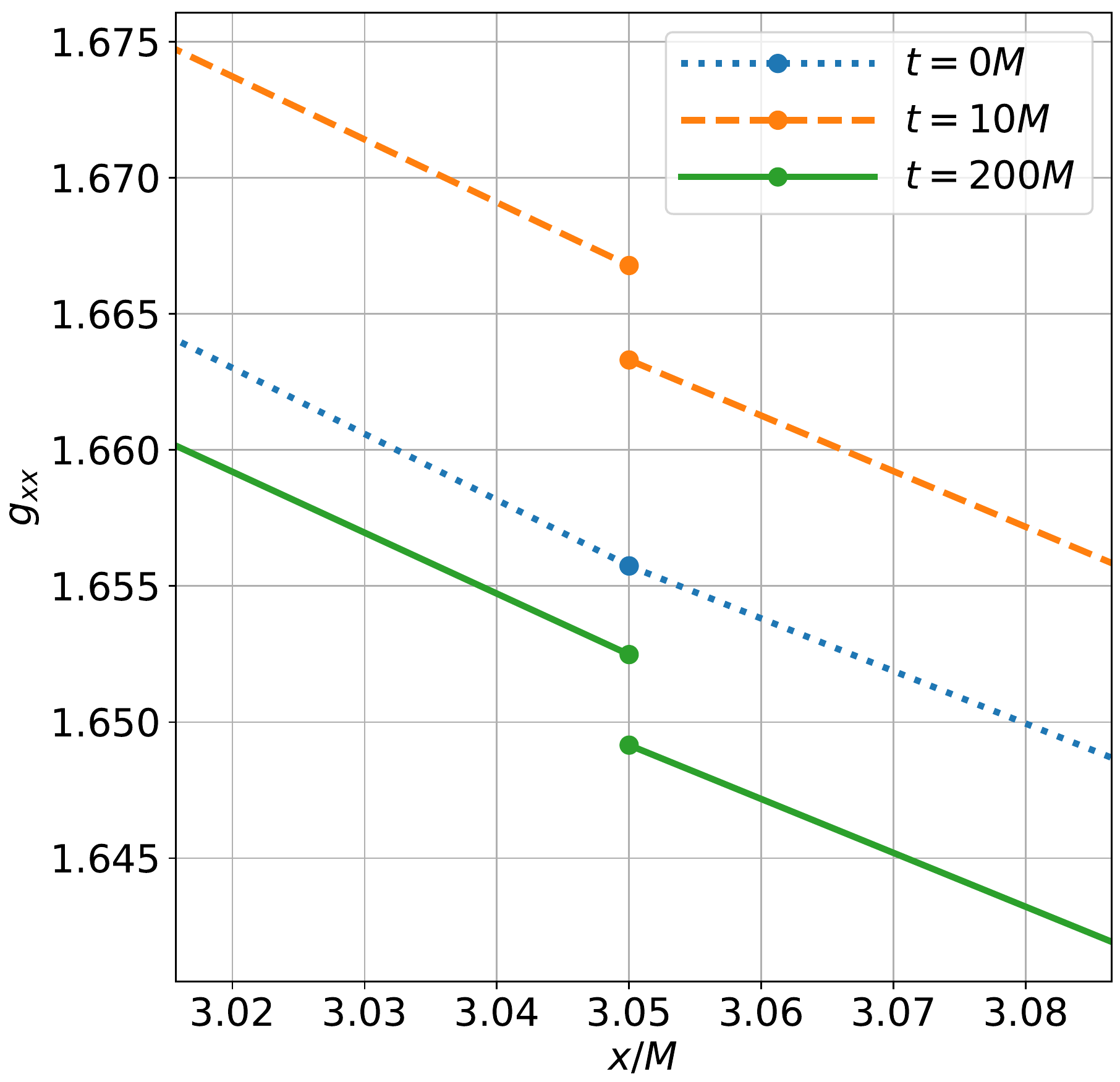}
\caption{
\label{KS_gxx_discont_zoom}
The plot shows the metric component $g_{xx}$ at different times, along a
portion of the $x$-axis near a domain boundary located at $x=3.05M$. The
initial data at $t=0M$ are continuous (dotted line). By $t=10M$ a
discontinuity (dashed line) has developed. The metric still rapidly evolves
for another $100M$ and then stabilizes around $t=200M$, while keeping the
size of the discontinuity roughly constant. Any further metric changes
after $t=200M$ are so small that they are practically indistinguishable
from the solid line.
}
\end{figure}
In Fig.~\ref{KS_gxx_discont_zoom} we plot the metric component $g_{xx}$ at
different times (for the $n=4$ case of Fig.~\ref{KS_stability}).
We have zoomed in onto a region close to a domain boundary
that is in the strong field region close to the horizon at $x=2M$.
We find that the metric rapidly evolves away from the
continuous Kerr-Schild initial data. During this rapid evolution
discontinuities develop across domain boundaries due to numerical errors.
These discontinuities persist throughout the evolution, but do not
negatively affect its stability or convergence, even though dynamic evolution
takes place. This indicates that such discontinuities in the physical metric
are not a problem for a DG method that uses numerical fluxes and
eigenvalues, which are computed from normals that are normalized with
respect to the flat metric.

\subsection{General relativistic hydrodynamics}

To treat neutron star matter we use the Valencia formulation~\cite{Banyuls97}. Matter is thus
described as a perfect fluid, where the stress-energy tensor is given by
\be
\label{eq:Tmunu}
  T^{\mu\nu} = \left(\rho + P \right) u^\mu u^\nu + P g^{\mu \nu} .
\ee
Here $\rho$ is the energy density, $P$ is the pressure,
and $u^\mu$ is the four-velocity.
The total energy density is written as
\be
\rho=\rho_0(1+\epsilon) ,
\ee
where $\rho_0$ is the rest-mass energy density,
and $\epsilon$ is the specific internal energy.
We express the four-velocity $u^\mu$ in terms of the three-velocity
given by
\be
v^{\mu} = u^{\mu}/W - n^{\mu}
\ee
and also introduce the Lorentz factor
\be
W = -n_{\mu} u^{\mu} = \alpha u^0.
\ee
Together $(\rho_0, \epsilon, Wv^i, P)$ are known as the primitive variables.

The matter equations follow from the conservation law for the
energy-momentum tensor and the conservation law for the baryon number.
In order to obtain flux conservative evolution equations of the
form~(\ref{Cons2}), one introduces the conserved variables
\ba
\label{D-def}
D    &=& \rho_0 W, \\
\label{tau-def}
\tau &=& \rho_0 h W^2 - P - \rho_0 W, \\
\label{Si-def}
S_i  &=& \rho_0 h W^2 v_i.
\ea
Here $D$ is rest-mass density, $\tau$ the internal energy density, $S_i$ the
momentum density as seen by Eulerian observers. The last two equations
also contain the specific enthalpy given by
\be
h = 1+\epsilon + P/\rho_0 .
\ee
The conserved variables are then
\be
u = \sqrt{\gamma}
\label{GRHDcons}
\left(
\begin{array}{ccc}
D    \\
\tau \\
S_l  \\
\end{array}
\right).
\ee
They satisfy Eq.~(\ref{Cons2}), with the flux vectors and sources given by
\be
\label{GRHDfluxes}
f^i = \sqrt{\gamma}
\left(
\begin{array}{ccc}
(\alpha v^i - \beta^i) D				\\
(\alpha v^i - \beta^i) \tau + \alpha P v^i		\\
(\alpha v^i - \beta^i) S_l + \alpha P \delta^{i}_l	\\
\end{array}
\right)
\ee
and
\be
\label{GRHDsources}
\frac{s}{\sqrt{\gamma}} =
\left(
\begin{array}{ccc}
0 \\
T^{00} (\beta^i \beta^j K_{ij} - \beta^i \partial_i\alpha) +
	T^{0i} (2\beta^j K_{ij} - \partial_i\alpha)
	+ T^{ij} K_{ij}	\\
T^{00}(\frac{\beta^i\beta^j}{2}\partial_l\gamma_{ij} - \alpha\partial_l\alpha)+
	T^{0i} \beta^j \partial_l\gamma_{ij}
	+ T^0_i \partial_l\beta^i
	+ \frac{T^{ij}}{2}\partial_l\gamma_{ij}	\\
\end{array}
\right).
\ee
The components of the stress-energy tensor appearing here can be expressed
in terms of the primitive variables as
\ba
T^{00} &=& (W^2 h\rho_0 - P)/\alpha^2, \\
T^{0i} &=& W h\rho_0 u^i/\alpha   + P\beta^i/\alpha^2, \\
T^{ij} &=& h \rho_0 u^i u^j + P (\gamma^{ij} - \beta^i\beta^j/\alpha^2), \\
T^0_i  &=& h \rho_0 W^2 v_i/\alpha,
\ea
where
\be
u^i = W v^i - W\frac{\beta^i}{\alpha}.
\ee

To close the evolution system, we have to specify an EoS for the fluid,
i.e. an equation of the form
\be
P = P(\rho_0, \epsilon)
\ee
that allows us to obtain the pressure for a given rest-mass energy density
and the specific internal energy, as well as the sound speed squared
$c^2_\mathrm{s}$.
If $c^2_\mathrm{s}<0$ or $c^2_\mathrm{s}>1$, we set it to zero. We also set it to zero if
$\rho_0=0$ or $h=0$.

As numerical flux we use the LLF flux of Eq.~(\ref{LLF_flux}). For this, we
need the eigenvalues of the characteristic modes given by~\cite{Banyuls97}
\ba
\lambda_1 &=& \alpha \frac{v^n (1-c^2_\mathrm{s}) + \sqrt{C^2}}{1-v^2 c^2_\mathrm{s}} - \beta^n, \\
\lambda_2 &=& \alpha \frac{v^n (1-c^2_\mathrm{s}) - \sqrt{C^2}}{1-v^2 c^2_\mathrm{s}} - \beta^n, \\
\lambda_3 &=& \lambda_4 = \lambda_5 = \alpha v^n - \beta^n,
\ea
where
$C^2 = c^2_\mathrm{s} (1-v^2)[ \gamma^{nn}(1-v^2 c^2_\mathrm{s})
                              - v^n v^n(1-c^2_\mathrm{s}) ]$,
$v^n = v^i n_i$, and $n_i$ is the normal to the interface,
normalized with respect to the flat metric.
At points where $1-v^2 c^2_\mathrm{s} = 0$ or $C^2<0$, we simply set
$\lambda_1=\lambda_2=0$.

\subsubsection{Converting conserved to primitive variables}

As already mentioned, we formulate the matter equations in the flux
conservative form of Eq.~(\ref{Cons2}) in terms of the conserved variables
in Eq.~(\ref{GRHDcons}). However, the flux vectors and sources in
Eqs.~(\ref{GRHDfluxes}) and (\ref{GRHDsources}) also depend on the primitive
variables $\rho_0$, $\epsilon$, $Wv^i$, $P$. Thus we need a way to compute
the primitive variables form the conserved variables. This is done with
the help of a root finder that we will describe next.
Note that we use $Wv^i$ as our primitive velocity variable instead of $v^i$.
The advantage is that $Wv^i$ is allowed to take any real value, while $v^i$
is bounded by the speed of light. The latter is inconvenient in numerical
calculations as numerical inaccuracies can often violate the light speed
bound.

The method we use closely follows the approach in appendix C
of~\cite{Galeazzi:2013mia}, i.e. we will try to find
\be
Wv := \sqrt{Wv^i Wv_i}
\ee
with the help of a root finder. This root is given by the zero of the
function
\be
f(Wv) = Wv - \frac{\sqrt{S_i S^i}}{D h(Wv)} .
\ee
Here, in order to find $h(Wv)$, we first need to compute the following:
\ba
W            &=& \sqrt{1 + (Wv)^2},  \\
\rho_0(Wv)   &=& \frac{D}{W},        \\
\epsilon(Wv) &=& W \frac{\tau}{D} - Wv\frac{\sqrt{S_i S^i}}{D}
                 + \frac{(Wv)^2}{1 + W}, \\
P(Wv)        &=& P(\rho_0(Wv), \epsilon(Wv)) \\
a(Wv)        &=& \frac{P(Wv)}{\rho_0(Wv) + \rho_0(Wv)\epsilon(Wv)}, \\
h(Wv)        &=& [1 + \epsilon(Wv)][1 + a(Wv)] .
\ea
Note that our implementation of the EoS $P(\rho_0, \epsilon)$ gracefully
handles cases where $\epsilon$ is slightly negative. Nevertheless if
$\epsilon(Wv) < 0$ we set it to zero when calculating $a(Wv)$ and $h(Wv)$.

After we have obtained the primitive variables, we calculate
$Z^i = S^i / (W h \rho_0)$ and $Z = \sqrt{Z^i Z_i}$.
According to Eq.~(\ref{Si-def}), we should have $Z^i = W v^i$. However, due
to numerical errors, the latter equality will only hold up to the accuracy
goal specified for the root finder (typically, the root finder has a
relative error of $10^{-10}$).
If $Z \leq Wv$, we accept this small discrepancy, but if
$Z > Wv$, we scale both $S_i$ and $Wv^i$ by a factor of $Wv/Z$.

\subsubsection{A positivity limiter for low density regions}


We use a strong stability preserving third order Runge-Kutta
scheme~\cite{Gottlieb2001} to
evolve the conserved variables. It is possible that the conserved variables
become unphysical after a Runge-Kutta substep due to numerical errors.
By unphysical, we mean points where the mass density $D$ or the energy
density $\tau$ is negative, or where $S > D + \tau$, with
$S=\sqrt{S_i S^i}$. If this happens,
it also becomes impossible to then find the primitive variables needed for
the next Runge-Kutta substep. To combat this problem we use so-called
positivity limiters after each substep. The idea of these limiters is
to scale each conserved variable $u$, that we desire to limit,
towards its node average $\bar{u}$ using
\be
\label{pos_limiter}
u \rightarrow \bar{u} + \theta_u \cdot (u - \bar{u}) .
\ee
Here $0 \leq \theta_u \leq 1$, and $u$ can be $D$, $\tau$ or $S_i$. For each
we try to find the maximum $\theta_u$, such that $u$ satisfies certain
criteria. For $D$, the criterion is $D \geq 10^{-12} \rho_{0,\mathrm{max}}$, where
$\rho_{0,\mathrm{max}}$ is the maximum mass density. For $\tau$, we simply demand
$\tau \geq 0$, while the $S_i$ criterion is $S < D + \tau$.
All three criteria have to hold at each point of the node.
Of course even with the
lowest allowed value of $\theta_u=0$, it is possible that some of the three
criteria are still not met at some points.
This occurs if $\bar{D} < 10^{-12} \rho_{0,\mathrm{max}}$ or $\bar{\tau} < 0$. In
this case we replace $\bar{D}$ or $\bar{\tau}$ in Eq.~(\ref{pos_limiter})
by these limits. If $\bar{S} > \bar{D} + \bar{\tau}$
we reduce the magnitude of the vector $S_i$ by a factor of
$(\bar{D} + \bar{\tau})/\bar{S}$ to meet
this criterion.
Notice that we do not use an artificial atmosphere as,
e.g.,~in~\cite{Font98b,Dimmelmeier02a,Baiotti04a,Yamamoto:2008js,
Thierfelder:2011yi,Galeazzi:2013mia,Rezzolla:2013,Baiotti:2016qnr,
Deppe:2021bhi}.
Rather the positivity limiters described above ensure that
$D \geq 10^{-12} \rho_{0,\mathrm{max}}$, $\tau \geq 0$,
and $S < D + \tau$.
In some sense that gives us an atmosphere as well, as $D$ can never drop
below this minimum. Yet, since in most cases scaling towards the average
suffices to satisfy all three criteria, we do not violate mass, energy or momentum conservation in most cases. And even in cases where we reset $D$,
$\tau$ or $S_i$ in some node, and thus violate conservation, we usually need
to modify only one of these conserved variables, while the usual artificial
atmosphere treatment would set $D$ to an atmosphere value and also zero both
$\tau$ and $S_i$, thus removing any velocity that the atmosphere naturally
might have had. As shown in~\cite{Poudel:2020fte}, resetting as little as
possible can be an advantage in simulations with orbiting stars when we wish
to accurately track lower density mass ejecta.

\subsubsection{Star surfaces}

Since the matter fields are not smooth across neutron star surfaces, we observe Gibbs
phenomena (i.e. high frequency noise) in the nodes that contain a piece of
the star surface. Here, we use a simple solution to this problem and apply
the filter of Eq.~(\ref{exp_filter}) to damp this noise after each full time
step. This filter changes the fields at every point by a typically small
amount. Nevertheless this can still cause trouble in low density regions by,
e.g., making $D$ or $\tau$ slightly negative or by violating
$S < D + \tau$. Thus after filtering we reapply the positivity
limiters discussed above. For the neutron star tests described below, we use
the filter parameters $\alpha_\mathrm{f}=36$ and $s=32$.

\subsubsection{Tests with single neutron stars}

To test our general relativistic hydrodynamics implementation, we
have performed simulations of a single neutron star for a fixed spacetime metric.
As already mentioned, we use units where $G=c=M_{\odot}=1$. To convert to
SI units, a dimensionless length has to be multiplied by $L_0 = 1476.6250$~m,
a time by $T_0 = 4.9254909\times10^{-6}$~s,
a mass by $M_{\odot}=1.9884099\times10^{30}$~kg,
and a mass density by $6.1758285\times10^{20}$~kg/m$^3$.

The first test starts with initial data for a static
Tolman-Oppenheimer-Volkoff (TOV) star with
a central density of $\rho_0 = 0.00128$. To setup the initial data,
we use a polytropic EoS, where pressure and specific internal energy are
given by
$P=\kappa\rho_0^{1+1/n}$ and
$\epsilon=n\kappa\rho_0^{1/n}$,
with $\kappa=100$ and $n=1$.
This results in star with a baryonic mass (i.e. rest-mass) of
$m_0 = 1.5061762 M_{\odot}$ and an ADM mass of $m=1.4001597 M_{\odot}$.
For the subsequent evolution we adopt a gamma-law EoS of
the form $P=\rho_0\epsilon/n$ with $n=1$.

We evolve this star on a single cubic patch with side length $32$. The patch
is centered on the star and covered by Cartesian coordinates. To better
resolve the star surface, where the matter fields are not smooth, we use
either 4, 5, or 6 levels of h-refinement, so that we end up with 4096,
32768, or 262144 leaf nodes. In each node, we use $5\times 5\times 5$
grid points. The star is then evolved for more than $5000 M_{\odot}$,
with a time step of $0.1$, $0.05$, or $0.025$.
\begin{figure}
\centering
\includegraphics[width=\linewidth]{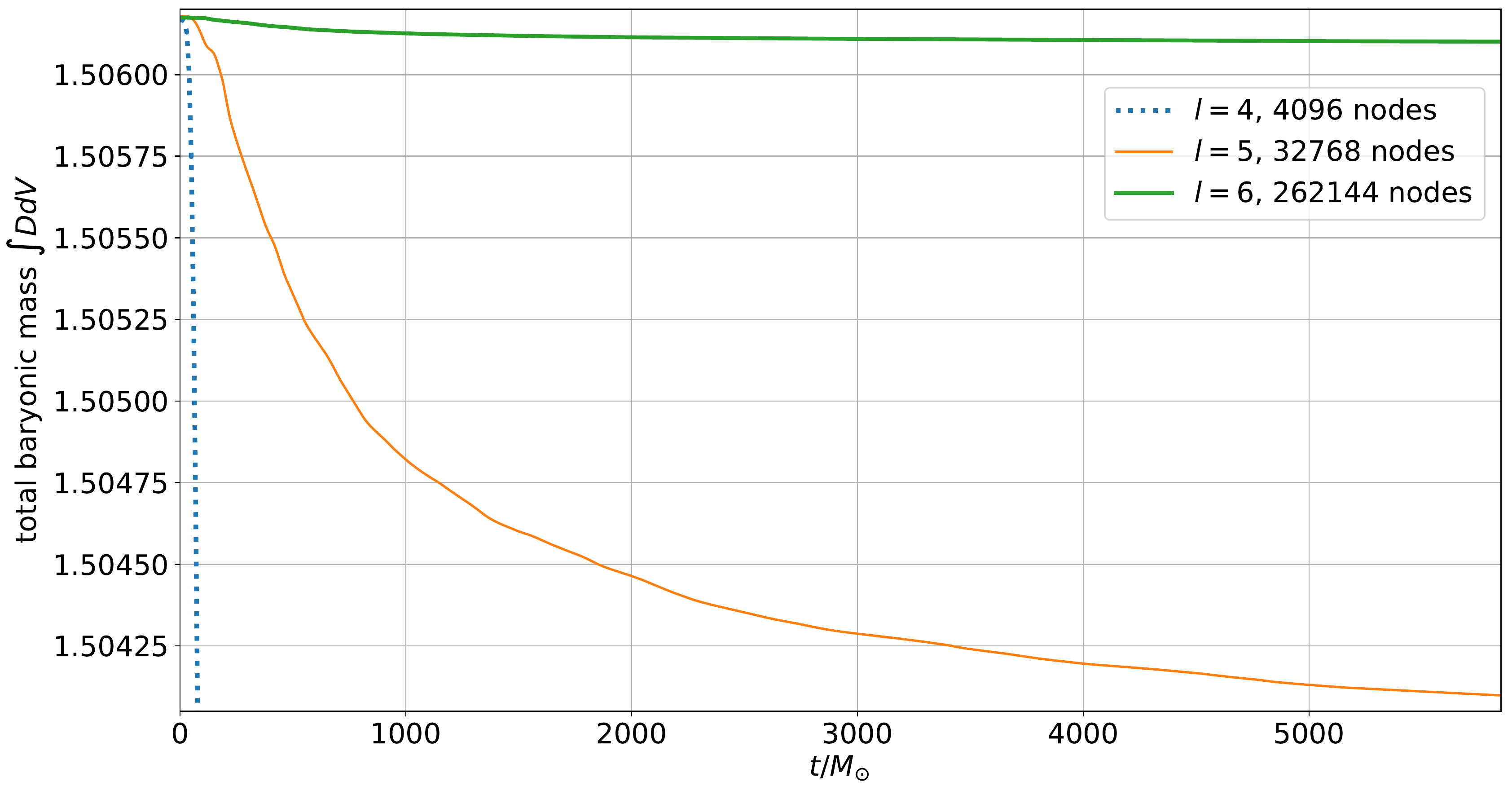}
\caption{
\label{TOV-D_VolInt}
The plot shows the total baryonic mass in our computational domain
versus time for $l=5$ and $l=6$ levels of h-refinement.
As one can see, mass conservation is quite good
at the highest resolution. Here $dV = \sqrt{\gamma}d^3x$.
}
\end{figure}
As we can see in Fig.~\ref{TOV-D_VolInt}, baryonic mass conservation
improves with increasing resolution. The reason why mass is not exactly
conserved is twofold. First, as already mentioned, our positivity limiters
are conservative only if the node average is above the limits we impose. Yet
this is not always the case, so that the limiter can cause conservation
violations. Second, the outer boundary is relatively close, so that mass can
escape from our numerical domain. Nevertheless baryonic mass conservation
improves with increasing resolution.

\begin{figure}
\centering
\includegraphics[width=\linewidth]{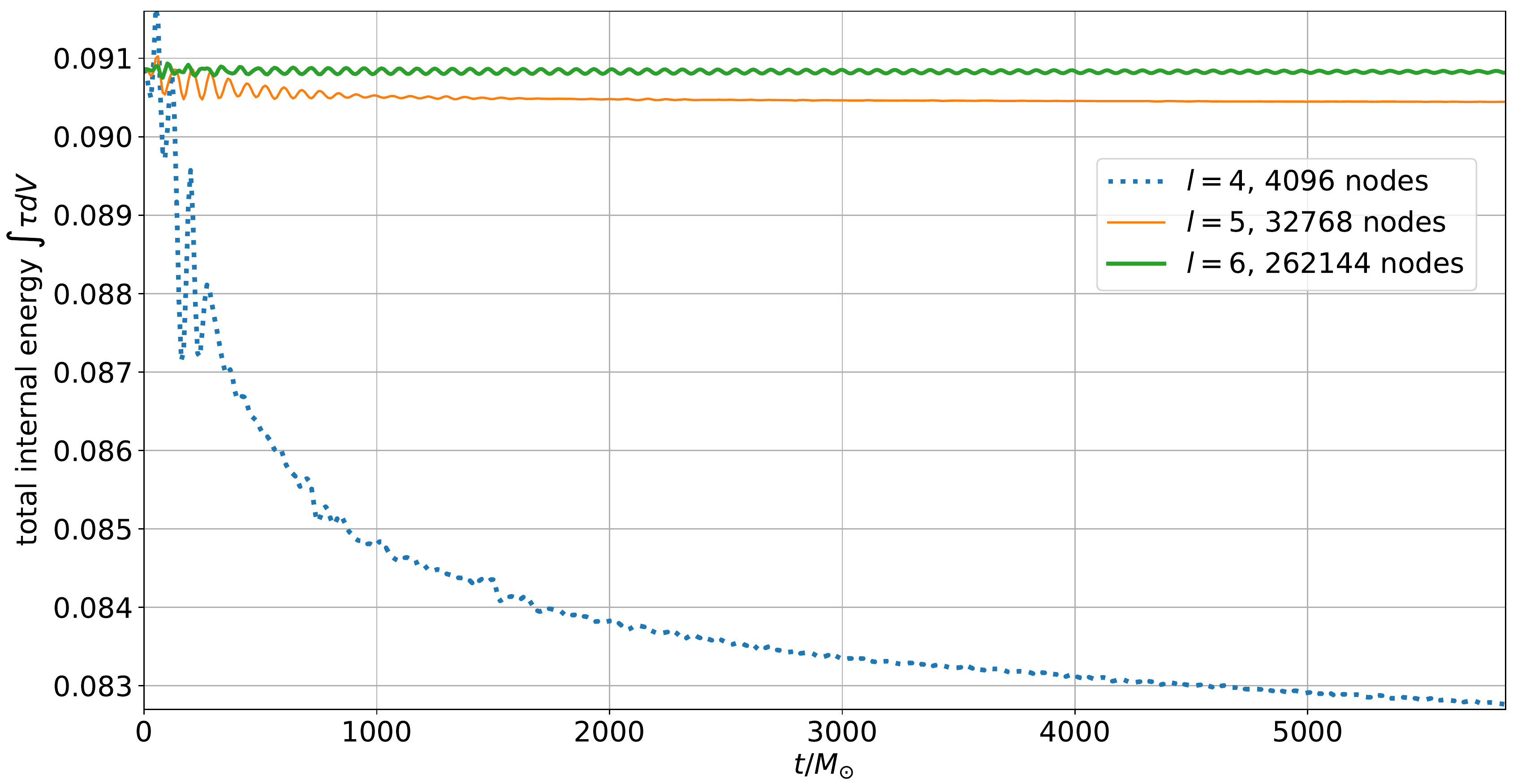}
\caption{
\label{TOV-Tau_VolInt}
The plot shows the total internal energy
versus time for $l=4$, $l=5$, and $l=6$ levels of h-refinement.
Since $\tau$ is not strictly conserved in general relativity, we can see
oscillations in it. For $l=5$ and $l=6$, the oscillation period of
$75 M_{\odot}$ is easily visible and agrees with the fundamental
oscillation frequency of the star.
}
\end{figure}
In Fig.~\ref{TOV-Tau_VolInt} we show the integral over the internal energy
density $\tau$.
This quantity is conserved in special relativity, but has a source
term in general relativity. Thus, it is not expected to be strictly conserved
during an evolution. In fact, for the two higher resolutions, one can clearly
see oscillations in it that are slowly damped out. The period of these
oscillations is about $75 M_{\odot}$ which corresponds to a frequency of
2.7~kHz, which is in agreement with the known fundamental oscillation
frequency of this star~\cite{Font:2001ew,McDermott1983}.
\begin{figure}
\centering
\includegraphics[width=\linewidth]{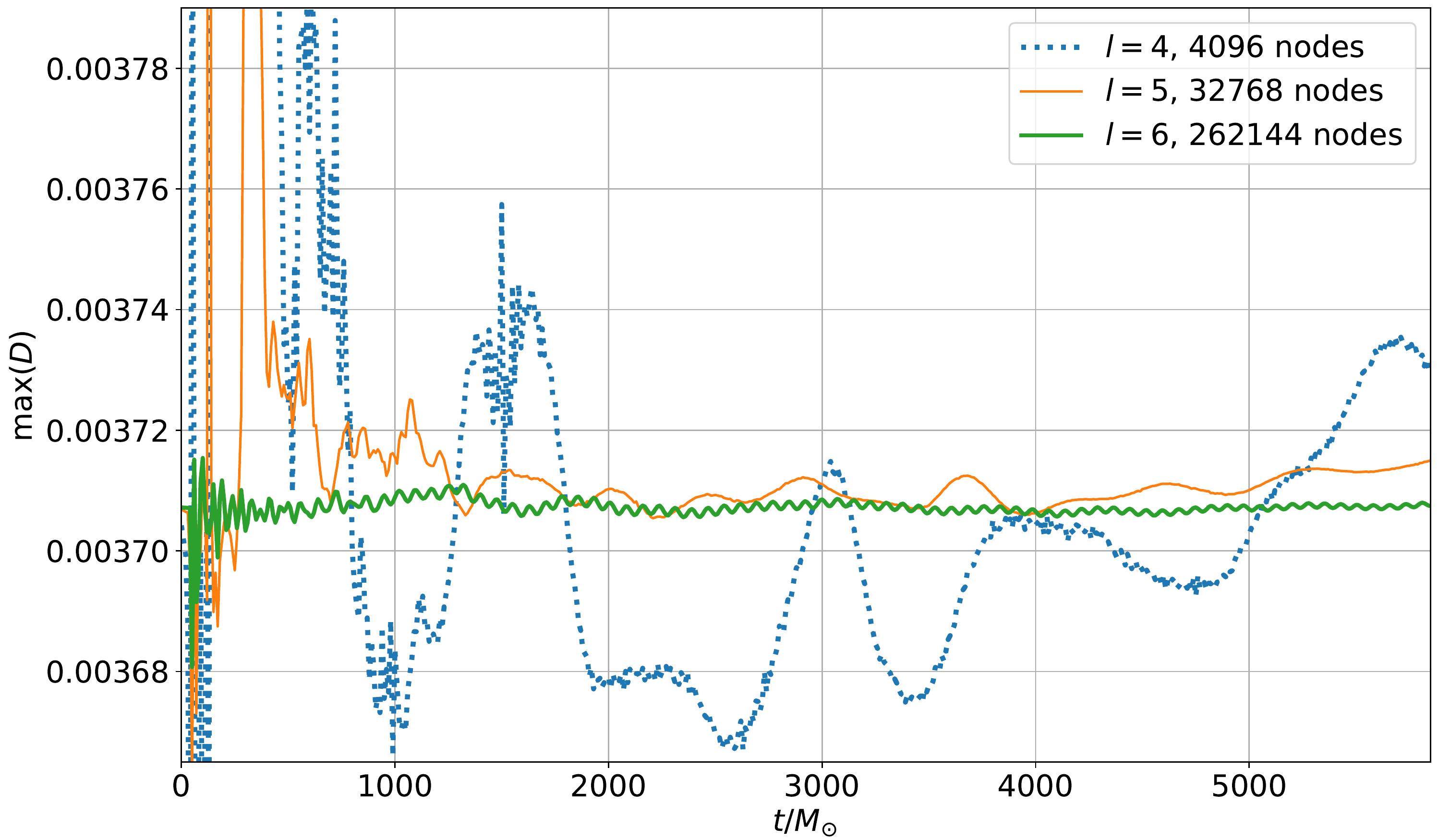}
\caption{
\label{TOV-D_max}
The plot shows the maximum of $D$
versus time for $l=4$, $l=5$, and $l=6$ levels of h-refinement.
Gibbs phenomena emanating from the star surface lead to noisy oscillations.
Only for the highest resolution, these oscillations clearly exhibit the
fundamental oscillation frequency of this star.
}
\end{figure}
These oscillations are also visible for the
highest resolution in Fig.~\ref{TOV-D_max}, which shows the maximum of $D$
versus time. For the lower two resolutions, however, these oscillations are
swamped by noise that is caused by Gibbs phenomena at the star surface.
The reason why oscillations due to Gibbs phenomena are more prominent in
Fig.~\ref{TOV-D_max} than in Figs.~\ref{TOV-D_VolInt} and
\ref{TOV-Tau_VolInt} is that the maximum of $D$ is determined at a single
point, while the integrals over $D$ and $\tau$ represent an average over the
entire domain that is less sensitive to Gibbs phenomena.
It is clear from Fig.~\ref{TOV-D_max} that if we are interested in values
at particular points, we need high resolution to get results where the
expected physical oscillations dominate over the oscillations due to Gibbs
phenomena. Nevertheless, our approach, that only uses positivity limiters
together with filters, is capable of stabilizing the evolution of the star
for all three resolutions.

The oscillations described so far originate purely from numerical errors. To
test the robustness of our approach, we have also evolved perturbed stars.
In this case, we use the same analytic TOV solution as above, but we add a
perturbation of the form
\be
\delta P = \lambda\cdot (P + \rho_0 + \rho_0\epsilon) \sin(\pi r/r_\mathrm{surf})
           Y^0_2(\theta,\varphi)
\ee
to the pressure. Here $(r,\theta,\varphi)$ are the standard spherical
coordinates, $r_\mathrm{surf}=8.1251439$ is the radius of the unperturbed star in
isotropic coordinates, and $Y^0_2(\theta,\varphi)$ is the $l=2$, $m=0$
spherical harmonic. We use the above polytropic EoS to then recalculate the
initial $\rho_0$ and $\epsilon$ from the perturbed pressure $P + \delta P$.
All metric variables are kept at their unperturbed TOV values. For the
simulations in this paper, we have used a fairly strong perturbation with
$\lambda=0.05$.

\begin{figure}
\centering
\includegraphics[width=\linewidth]{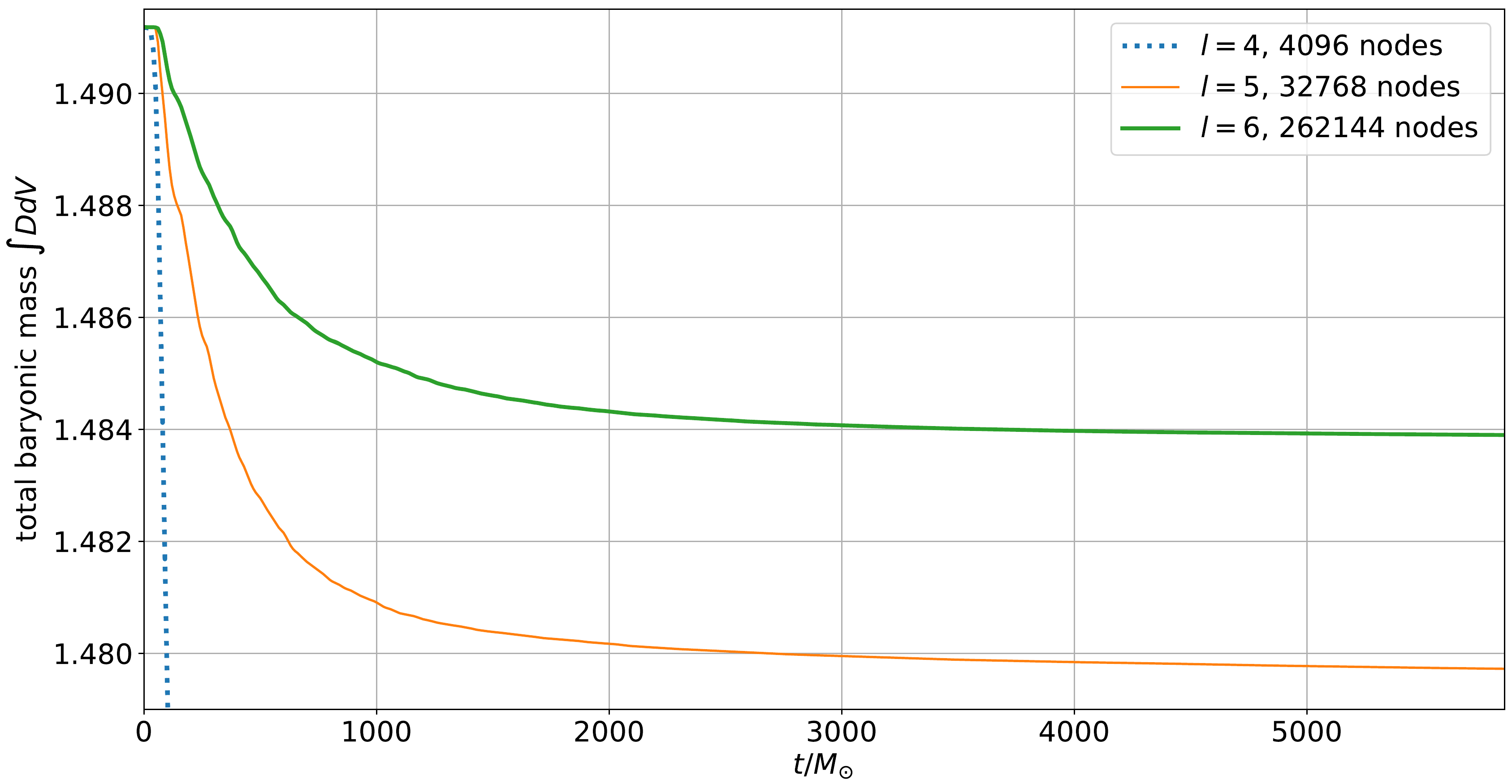}
\caption{
\label{pTOV-D_VolInt}
The total baryonic mass for the perturbed star versus time for
$l=5$ and $l=6$ levels of h-refinement.
Strong star pulsations cause material to leave through the outer
boundary and are thus responsible for the initial drop in the mass.
}
\end{figure}
\begin{figure}
\centering
\includegraphics[width=\linewidth]{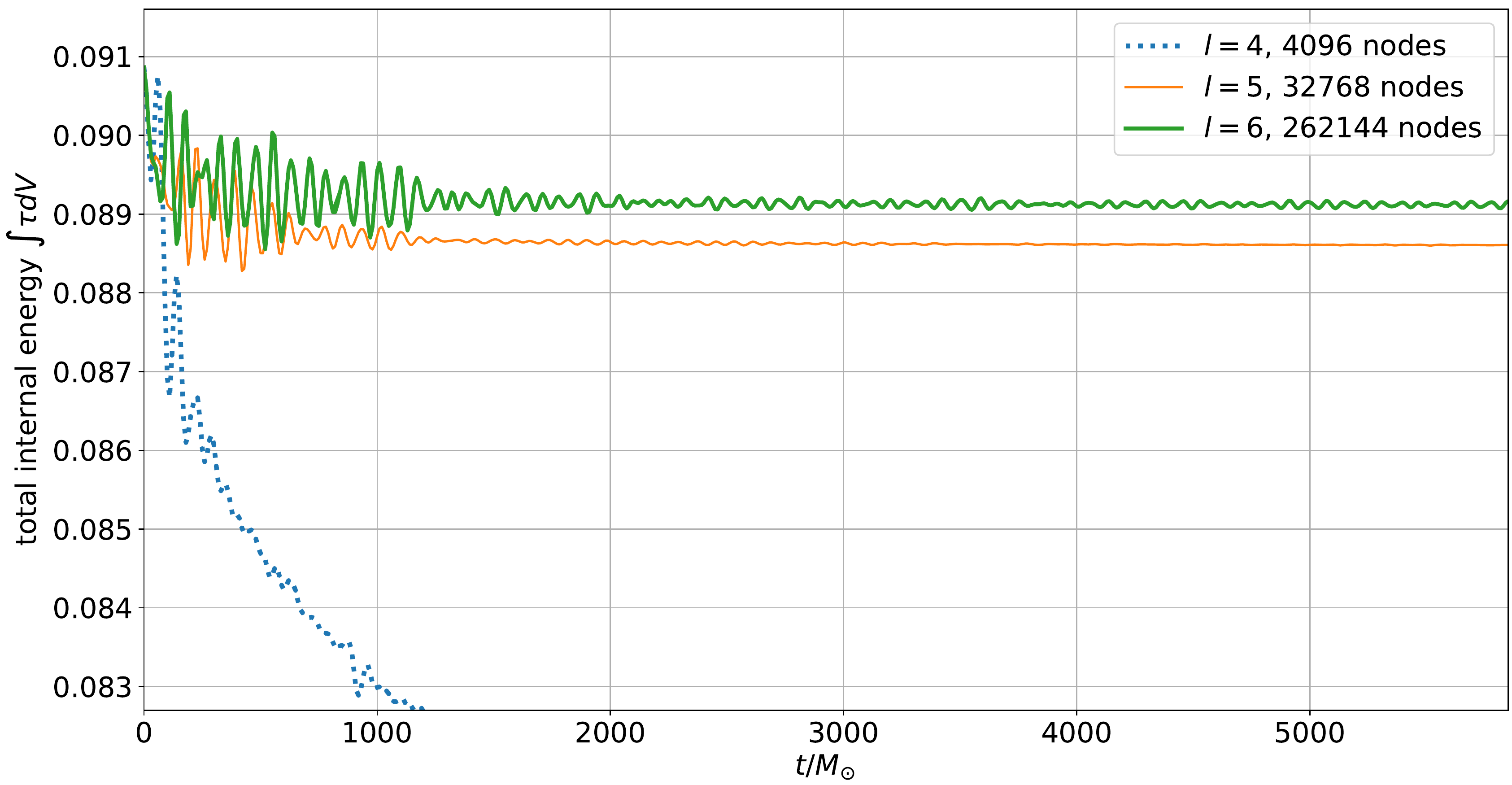}
\caption{
\label{pTOV-Tau_VolInt}
The total internal energy for the perturbed star. Due to the strength of
the perturbation the oscillation amplitude is much larger than for the
unperturbed star in Fig.~\ref{TOV-Tau_VolInt}.
}
\end{figure}
\begin{figure}
\centering
\includegraphics[width=\linewidth]{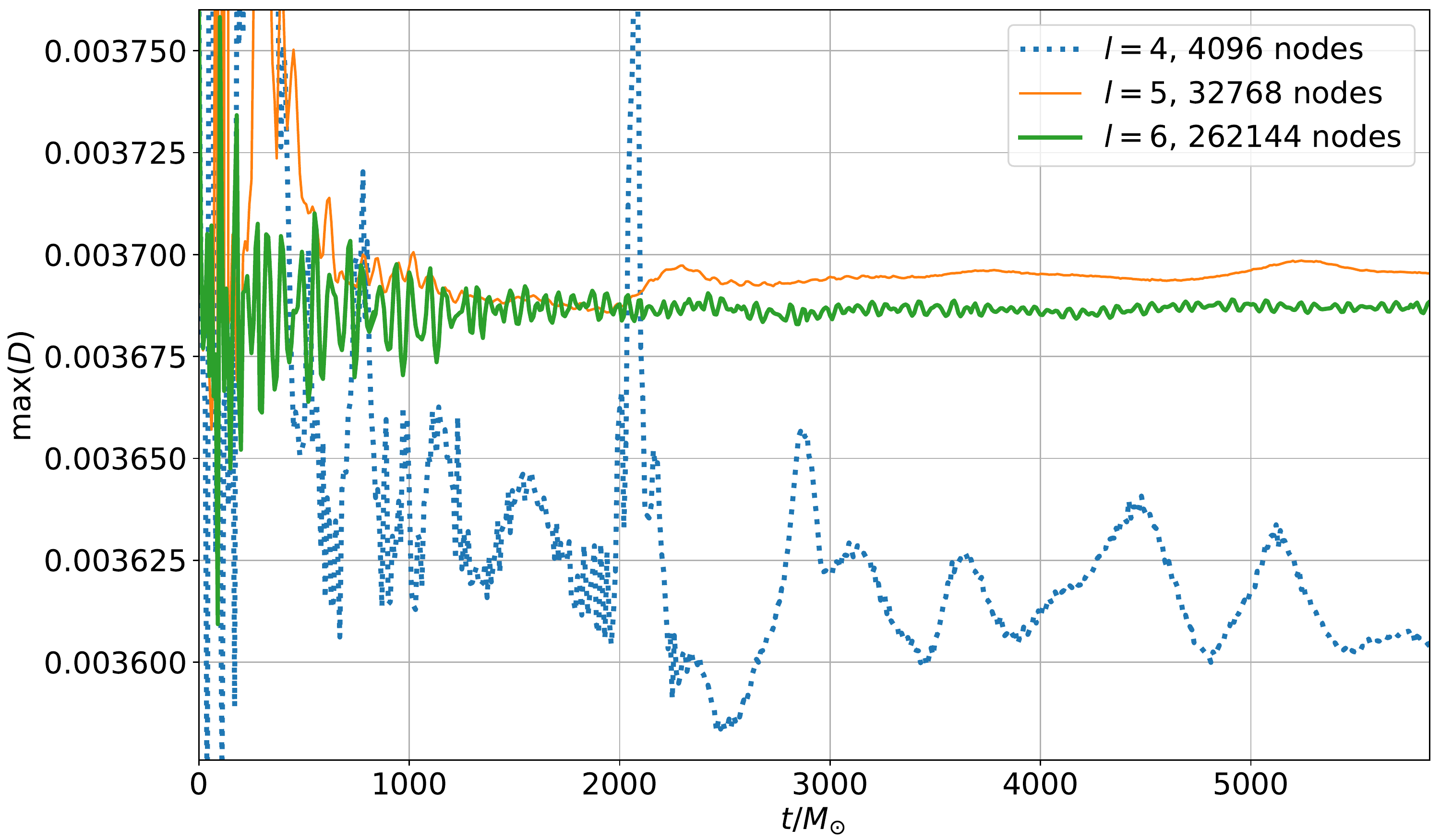}
\caption{
\label{pTOV-D_max}
The maximum $D$ for the perturbed star is qualitatively similar to the
unperturbed case, but the oscillation amplitudes are larger.
}
\end{figure}

In Figs.~\ref{pTOV-D_VolInt}, \ref{pTOV-Tau_VolInt} and \ref{pTOV-D_max},
we show the total mass, the total internal energy, and the maximum of the
density $D$ for the perturbed star.
Since the perturbation is relatively strong, the star oscillations are
now much bigger, so that the oscillations in the total internal energy
are now much larger. This is clearly visible for the two higher
resolutions ($l=5$, $l=6$) in Fig.~\ref{pTOV-Tau_VolInt}.
In fact, the star pulsations are now so strong that much more material leaves
through the outer boundary. This leads to the initial drop in the mass seen
in Fig.~\ref{pTOV-D_VolInt}.
The maximum of $D$ also oscillates stronger,
but as for the unperturbed case, the expected oscillation frequency
is only readily discernible at the highest resolution ($l=6$) in
Fig.~\ref{pTOV-D_max}.

When comparing the oscillations in the internal energies
in Figs.~\ref{pTOV-Tau_VolInt} and \ref{TOV-Tau_VolInt},
for both perturbed and unperturbed stars, we can see that in both cases
the star oscillations are more strongly damped for low resolutions.

The main takeaway is thus that our approach is robust since it still works
for strongly perturbed stars. We have also seen that, at the lowest
resolution, oscillations due to Gibbs phenomena can easily dominate the
expected physical oscillations. Such Gibbs phenomena will become only worse
once we have to deal with true shocks, e.g. if two stars collide.
We thus expect to need additional limiters once we have to deal with
shocks.

We also wish to comment on the work in~\cite{Deppe:2021bhi} where single
neutron stars are simulated using a DG method together with various limiters
(such as e.g. WENO, HWENO, or Krivodonova), and also using a hybrid scheme,
that switches to finite differences (FD) in non-smooth regions (e.g. near
the star surfaces). The main result of~\cite{Deppe:2021bhi} is that their
hybrid DG-FD scheme works better than any of the many limiters tested, and
that in fact the evolution of a single neutron star failed with many of the
limiters tested. Since our new positivity limiter is not expected to be
sufficient to deal with true shocks, using such a hybrid DG-FD scheme may
very well be the best way forward. However, it is possible we are at least
able to obtain stable evolutions with our limiter if we combine it with an
additional limiter that deals with shocks. In the next section we will
therefore test several limiters that are designed to treat shocks.


\subsubsection{Limiters for the treatment of shocks}

Since general relativistic hydrodynamics allows for the development of
shocks in the fluid, we need to be prepared to deal with them.
A general way to handle spurious oscillations due to Gibbs phenomena,
occurring in these situations, is to apply limiters to the hydrodynamic
fields. We try out two types of limiters in this paper. The first is
the so-called total variation bounded minmod or minmodB
slope limiter, which has been developed, demonstrated and utilized in
multiple articles, such as~\cite{Shu:1987a, Cockburn:1989a, Cockburn1998a, Schaal:2015ila},
including methods compatible with the DG evolution scheme. We follow closely the
formalism in~\cite{Schaal:2015ila} and apply the minmodB limiter
to the conserved variables.
The other one is the limiter proposed by Moe, Rossmanith and Seal
in~\cite{Moe:2015a}, dubbed henceforth as the MRS limiter.
In this work, we apply the MRS limiter to either the conserved
variables [MRS(cons.)] or the primitive variables [MRS(prim.)].
The case of MRS(cons.) is straightforward, as we can directly apply
the limiter to the variables we actually evolve. However, in case
of MRS(prim.), a problem arises since we first have to recover the
primitive variables from the evolved conserved variables, which
can fail if, e.g., the momentum density is too high. To address this,
we perform a procedure of prelimiting similar to what is described
in~\cite{Hebert:2018xbk}, to a copy of the conserved variables.
Through this prelimiting, we ensure that the strong
condition $S_i S^i < \tau (\tau + 2D)$ holds for this copy of
conserved variables.
Once we have calculated the primitive variables from the 
prelimited copy of conserved
variables, we compute the rescaling factor $\theta_i$ for the
MRS limiter, as described in~\cite{Moe:2015a}, using the
primitive variables $\rho_0$, $Wv^i$, and $P$.
However, after we have obtained this $\theta_i$,
we apply it to rescale the original non-prelimited conserved variables
that we are evolving, which is then our actual limiting procedure.

To test how well \nmesh~handles shocks, we implement
test cases in both 1D and 2D, where we have an initial
discontinuity in density and pressure, as in a
Riemann problem. We then evolve this initial discontinuity
using the full general relativistic hydrodynamic evolution
system of equations on a fixed Minkowski metric. The mesh is
composed of adjacent Cartesian domains. For these tests, the time
step was set to be $\Delta t = \Delta x_\mathrm{min} / 4$.


\begin{figure}
  \includegraphics[width=\linewidth]{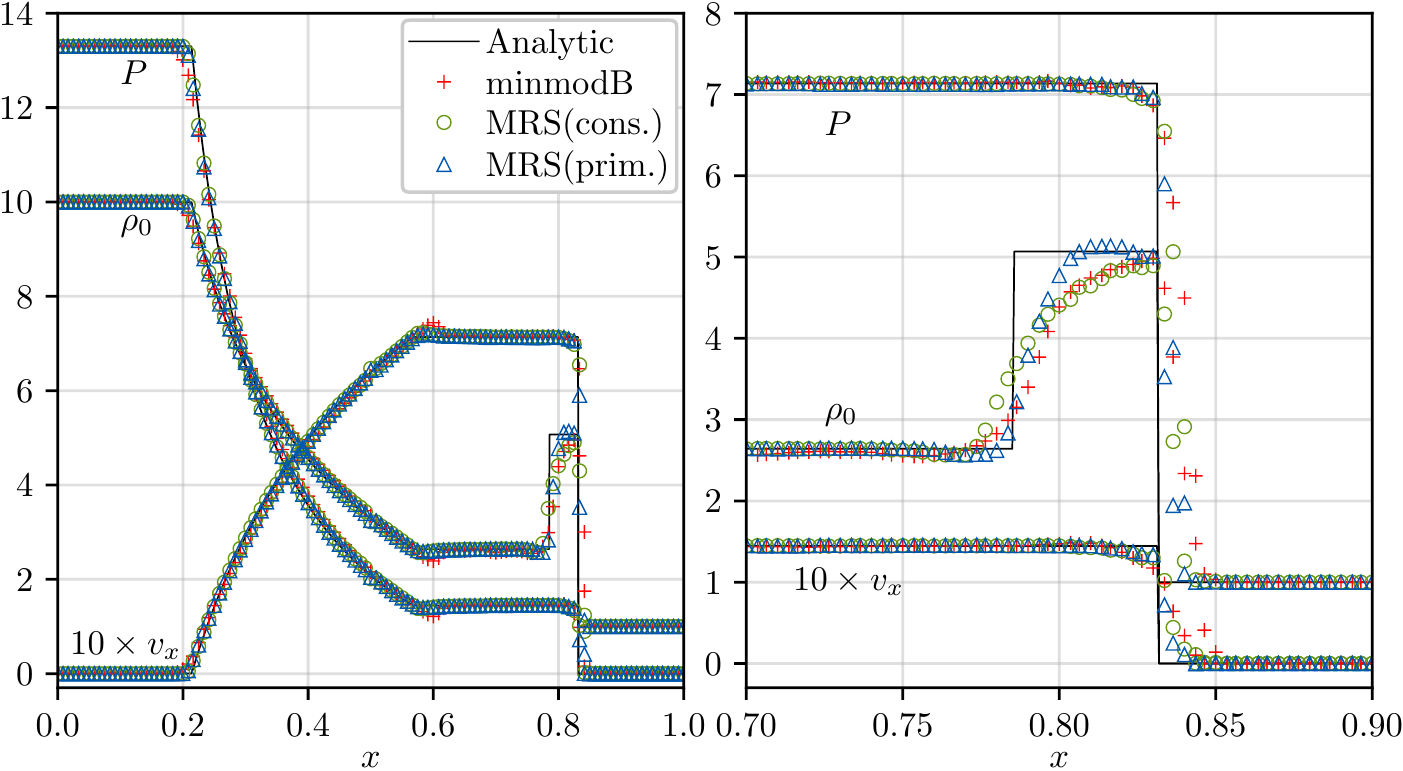}
\caption{\label{blastwav1d}
The plots show blast wave profiles for pressure $P$,
rest-mass density $\rho_0$ and speed $v_x$ (times 10)
at $t = 0.4$ after evolving the initial shocks in $P$ and $\rho_0$.
There are 200 domains, with 4 points per domain.
We show results for minmodB, MRS(cons.), and MRS(prim.) limiters.
The left plot shows the whole domain, while the right one
focuses on the contact discontinuity and shock fronts.
The legend on the left plot also holds for the right plot.
}
\end{figure}

\textit{1D Test:} We use the special relativistic blast wave
test from~\cite{Marti99}, and also use the analytic solution
code from the same article to compare with the numerical result from
\nmesh. The initial data in this case is such that we have
two different values on the left and right halves of
the mesh, for the primitive variables $\rho_0$ and $P$.
The values of the primitive variables $\rho_0$, $P$, and $v_x$
are as stated in Table~\ref{SRHD-1D-ID}.

\begin{table}[h!]
\caption{\label{SRHD-1D-ID}Initial data for 1D special
relativistic blast wave for primitive variables
$(\rho_0, P, v_x)$.}
\centering
\begin{tabular}{ |r| c|}
\hline
left,  $x\leq 0.5$ & (10.00, 13.33, 0.00) \\
right, $x > 0.5$   & (1.00, 0.00, 0.00) \\
\hline
\end{tabular}
\end{table}

In Fig.~\ref{blastwav1d} we show the profiles for the primitive variables
over the entire mesh (left plot), as well as the contact discontinuity and
the shock front (right plot), after evolving the initial data to
time $t = 0.4$.
The plots contain the numerical results obtained from \nmesh~as well as the
analytic solution from~\cite{Marti99}. For the numerical results,
we have used 200 adjacent Cartesian domains along the $x$-axis, with
4 points in each domain. For minmodB, referring to the formalism
in~\cite{Schaal:2015ila}, we set $\beta = 0.6$ and
$\alpha_\mathrm{lim} := \tilde{M} = 5$. For MRS(cons.) and MRS(prim.),
we set the $\alpha$ from~\cite{Moe:2015a} to
$\alpha = \alpha_\mathrm{lim} L^{3/2}$ with $\alpha_\mathrm{lim} = 25$,
where $L$ is the size of the node.
While the exact meaning of $\alpha_\mathrm{lim}$ is different in minmodB and MRS,
in both cases lower $\alpha_\mathrm{lim}$ makes the limiter more aggressive.
From the plots, it appears that the result with MRS(prim.) adheres closest
to the analytic result, whereas the one with minmodB seems to
deviate the most from it. This is true for the plot on the left,
that shows the behavior across the entire mesh, but is clearer
from the plot on the right, that focuses on the problematic
region of the contact discontinuity and the shock front.


\textit{2D Test:} The 2D test we perform is an extension of the
1D test Riemann problem, that can be found in~\cite{Zhao:2013a}.
The initial data in the primitive variables $(\rho_0, P, v_x, v_y)$
for the 2D test over the mesh is as stated in Table~\ref{SRHD-2D-ID}.

\begin{table}[h!]
\caption{\label{SRHD-2D-ID}Initial data for 2D special
relativistic blast wave for primitive variables
$(\rho_0, P, v_x, v_y)$.}
\centering
\begin{tabular}{ |c| c c|}
\hline
& $x<0$ & $x \geq 0$ \\
\hline
$y \geq 0$ & (0.1, 1, 0.7, 0) & (0.03515, 0.163, 0, 0) \\
$y<0$ & (0.5, 1, 0, 0) & (0.1, 1, 0, 0.7) \\
\hline
\end{tabular}
\end{table}

\begin{figure}
  \includegraphics[width=\linewidth]{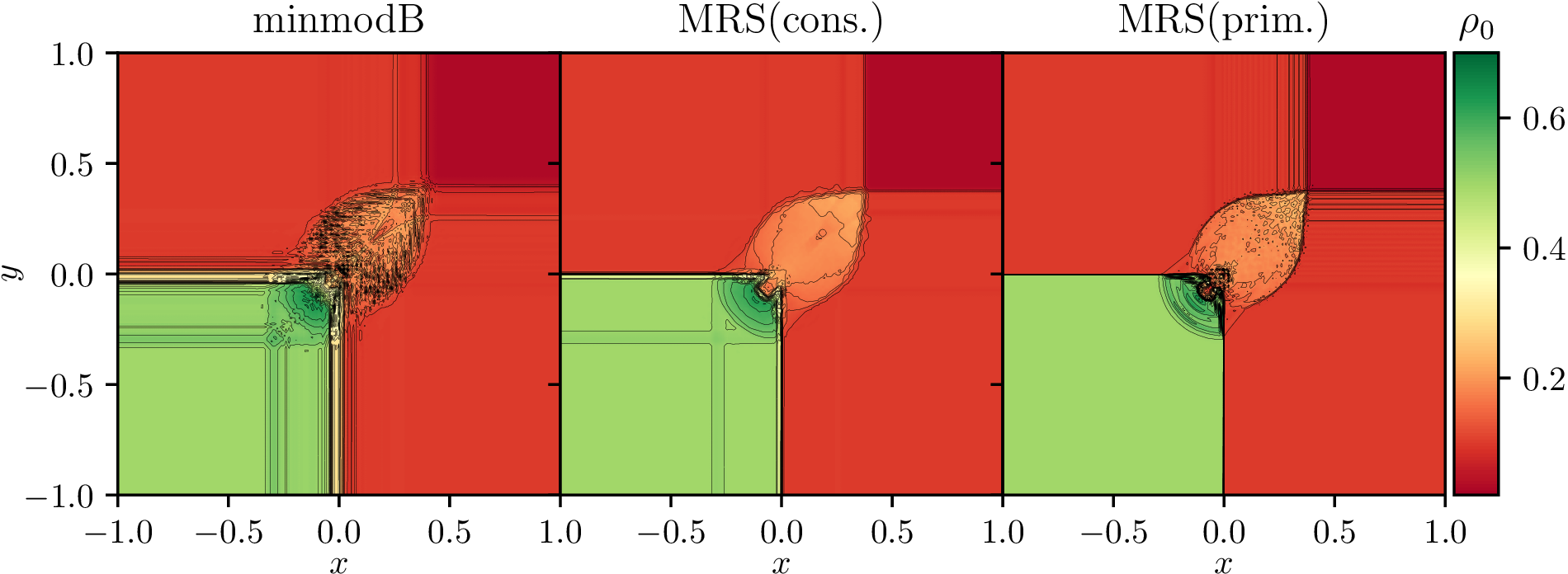}
\caption{\label{blastwav2d}
Plots showing rest-mass density profile of 2D blast wave
at $t = 0.4$ after evolving the initial discontinuity with minmodB,
MRS(cons.) and MRS(prim.) limiters in colormap and 30 contour
lines spaced evenly between 0.01 and 0.695.
}
\end{figure}
The numerical results obtained from the three different limiter
choices from \nmesh~are shown in Fig.~\ref{blastwav2d} at time
$t = 0.4$, after evolving the initial data.
We have only plotted the results for $\rho_0$, as it is
arguably the most problematic case. We compare our
results with those of Zhao and Tang in~\cite{Zhao:2013a}
and Bugner in~\cite{Bugner2018a},
while noting that Zhao and Tang have used a finite element
DG method with WENO and a special relativistic hydrodynamic
system of evolution equations and Bugner used a
DG method with WENO and fully general relativistic hydrodynamic
system of equations, while \nmesh~uses DG with the
minmodB and MRS limiters and the fully general relativistic hydrodynamic
system of equations. In our runs here, the mesh
is composed of $100 \times 100$ Cartesian domains, with 4 points,
i.e, we have $4 \times 4$ points in each domain
along each direction. Again, for minmodB,
we use $\beta = 0.6$ and $\alpha_\mathrm{lim} = 5$. For MRS(cons.) and MRS(prim.),
we set the parameter $\alpha_\mathrm{lim} = 25$. Also, we use our new positivity
limiter to control $S_i$ according to Eq.~\ref{pos_limiter}.

Once again, we see
that both MRS(cons.) and MRS(prim.) fare better than minmodB.
However, in this case, we cannot draw a clear conclusion
as to which one out of MRS(cons.) and MRS(prim.) yields
a better result. MRS(cons.) seems to provide overall
better results than MRS(prim.), except for the left bottom
region, where MRS(prim.) seems to be better at handling the
high density region and the so-called ``mushroom cloud"
structure around position (0,0). However, overall, the MRS limiter
cases are in reasonably good agreement with the results
found in~\cite{Bugner2018a,Zhao:2013a}.

\subsubsection{Single TOV star with MRS limiter}

Since limiting the conserved variables with the MRS scheme was successful
in our shock tube tests, we wanted to know how this limiter would influence
the neutron star simulations presented above. As a test we have repeated the runs for
the unperturbed TOV star, but this time with the MRS(cons.) limiter turned on with
$\alpha_\mathrm{lim}=25$. Note that the positivity limiters are still needed to deal
with low density regions. Thus we use both the MRS limiter and the
positivity limiters described above, with the MRS limiter running
immediately before the positivity limiters.
We find that the total baryonic mass and internal energy are almost the same
with and without the MRS limiter in the sense that the corresponding plots look
almost the same as in Figs.~\ref{TOV-D_VolInt} and \ref{TOV-Tau_VolInt},
even when the MRS limiter is turned on. The biggest difference occurs when
we plot the maximum of $D$.
\begin{figure}
\centering
\includegraphics[width=\linewidth]{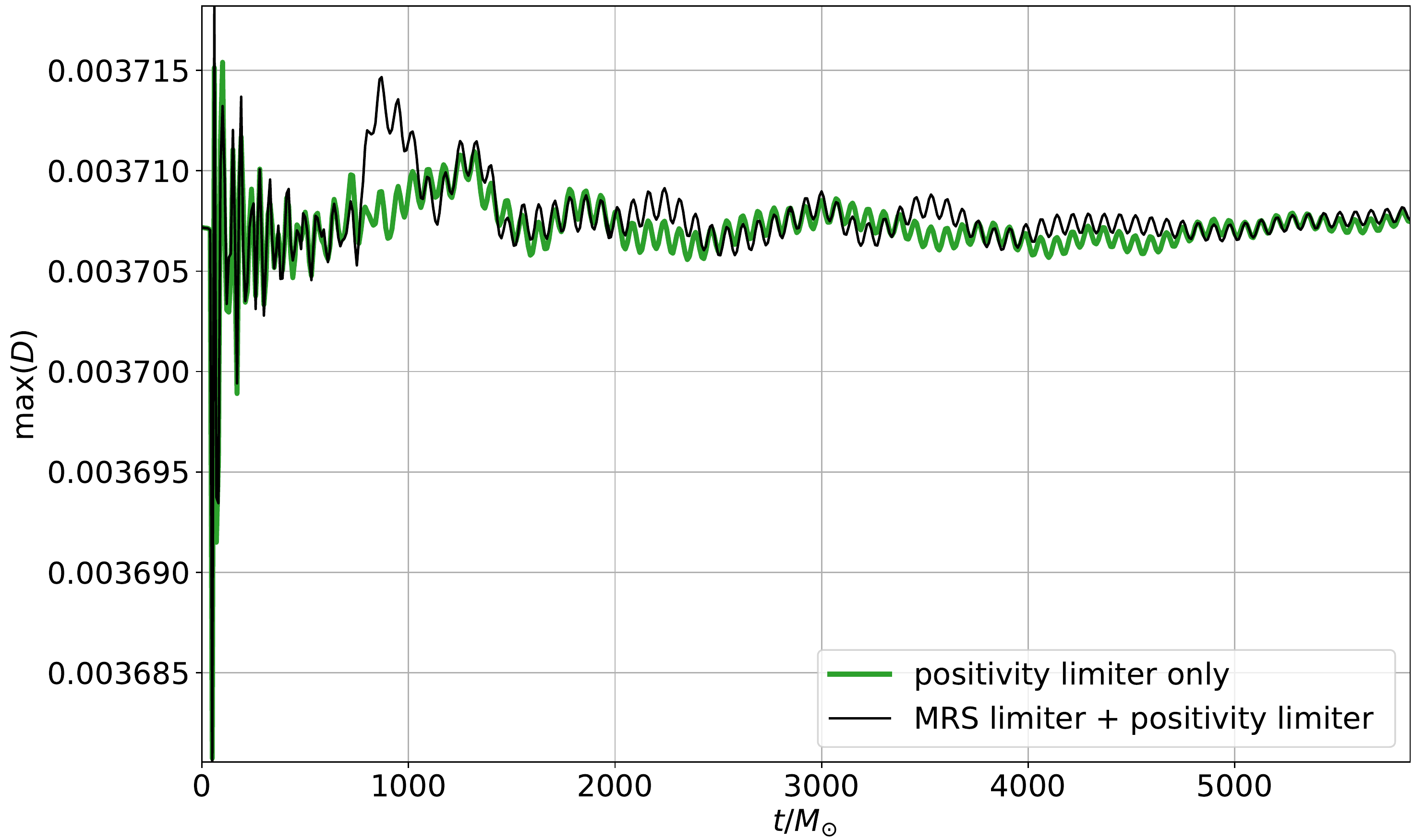}
\caption{
\label{TOV_MRS-D_max}
The plot shows the maximum of $D$ versus time for
six levels of h-refinement with (thin black line) and without (thick green line)
the MRS limiter turned on.
}
\end{figure}
In Fig.~\ref{TOV_MRS-D_max} we show the oscillations in the maximum
of $D$ for the high resolution with and without the MRS limiter.
In both cases we see the expected star oscillations, but there are also
longer term drifts in the maximum of $D$. With the MRS limiter this drift is
a bit different and arguably slightly more pronounced. Nevertheless, the
MRS limiter does not lead to big changes, which is not too surprising, since
the fluid in a single star does not contain any shock fronts. Yet, this run
demonstrates that the MRS limiter does not cause any stability problems
when added to our previous method.

\section{Discussion}
\label{discussion}

In this article, we have presented all the numerical and computational
methods used in our new \nmesh~program to evolve systems of hyperbolic
equations. The principal scheme we use for spatial discretization is a DG
method. This is then coupled with a Runge-Kutta time integrator to be able
to evolve in time. The DG method can easily deal with many domains. We use
this to introduce many patches, which can be adaptively refined by splitting
them into eight child domains (see e.g. Figs.~\ref{CubSphPatches}
and \ref{TOVref}), as often
as desired. This AMR scheme is then parallelized by distributing the
resulting many domains among all available compute cores. For the neutron
star test cases shown in Fig.~\ref{strongscaling} this approach achieves
good strong scaling. As explained in
Sec.~\ref{parallelization_strategy}, an advantage of DG methods is that
they result in less communication overhead than traditional finite
difference or finite volume methods. In~\cite{Deppe:2021bhi}, a similar DG
method is used, albeit with one crucial difference. To derive our DG method
we integrate using coordinate volume elements, and thus do not include the
physical metric. This leads to a simplification of the method where one does
not have to worry about possible discontinuities in the physical metric
or the normal vector across domain boundaries.

We have also carried out simulations of scalar fields and black holes
to test the convergence of our new program. Since in this case all evolved
fields are smooth, we expect exponential convergence when the
number of grid points is increased. Our simulation results conform to this
expectation. We find that both, the upwind and LLF fluxes perform equally
well, in all cases tested.

A much more complicated case is the evolution of neutron stars, since in
this case, the matter fields are not smooth across the star surface. An
additional problem arises from the fact that at each Runge-Kutta substep we
have to calculate the primitive variables from the evolved conserved
variables. The latter can easily fail in low density regions (such as the
vacuum outside the star), where numerical errors can cause the conserved
variables to become unphysical in the sense that the mass or internal energy
densities can become negative, or the momentum density can be become too
high. To address this problem, we have developed a new positivity limiter
that attempts to reset these variables by scaling toward their node averages
in case of trouble. If we use our positivity limiter together with the
exponential filters described before, we can stably evolve single neutron
stars. These stable evolutions are possible without any extra ingredients,
such as an artificial low density atmosphere or additional limiters (like
the minmodB or MRS limiters described above), that have been used in other
works. We believe that our positivity limiter is an important step, because
the more general limiters like minmodB or MRS are really designed to deal
with shocks and thus do not help in low density regions near star surfaces.
Nevertheless, something like these general limiters is still needed to deal
with shocks. As we have shown above, the general limiters can be used in
combination with the positivity limiter.

We thus have all necessary ingredients to perform simulations of binary
neutron stars and black holes, which is what we plan to do in the future
with the \nmesh~program.

\ack

It is a pleasure to thank Bernd Br\"ugmann for helpful discussions.
This work was supported by NSF grants PHY-1707227, PHY-2011729, and
PHY-2136036.
We also acknowledge usage of computer time
on the HPC cluster KoKo at Florida Atlantic University,
on the Dutch National supercomputer Cartesius, as well as
on Bridges-2 and Expanse under XSEDE allocation TG-PHY220018.


\appendix

\section{About the flat metric in Gau{\ss}'s theorem}
\label{Gauss_theorem}

In Eq.~(\ref{int_by_parts}) we have used Gau{\ss}'s theorem in the form
\be
\label{Gauss_law_flat}
\int \partial_i f^i d^3x = \oint f^i n_i d\bar{A}
\ee
where $n_i$ was normalized with respect to the flat metric, expressed as
$\delta_{ij}$ in the global Cartesian-like coordinates $x^i$ that cover all
our domains. For example, on $x^{\bar{3}}=1$ we have
\be
n_i = \frac{1}{\sqrt{\bar{\gamma}^{\bar{3}\bar{3}}}}
      \frac{\partial x^{\bar{3}}}{\partial x^i}
\ee
and
\be
\bar{\gamma}^{\bar{3}\bar{3}} =
\frac{\partial x^{\bar{3}}}{\partial x^i}
\frac{\partial x^{\bar{3}}}{\partial x^j}
\delta^{ij} .
\ee
Thus we find
\be
\label{n_i_dAbar}
n_i d\bar{A} = \frac{1}{\sqrt{\bar{\gamma}^{\bar{3}\bar{3}}}}
               \frac{\partial x^{\bar{3}}}{\partial x^i} d\bar{A}
             = \frac{\partial x^{\bar{3}}}{\partial x^i}
               \frac{J}{\sqrt{^{(2)}\bar{\gamma}}} d\bar{A}
             = \frac{\partial x^{\bar{3}}}{\partial x^i}
               J dx^{\bar{1}} dx^{\bar{2}} ,
\ee
where in the last two steps we have used Eqs.~(\ref{sqrtdiag_gammaii_flat})
and (\ref{dbarA_example}). We see that the flat metric pieces all cancel,
and thus do not influence the surface integral.

The analog of Eq.~(\ref{sqrtdiag_gammaii_flat}) for the physical
metric (denoted by $\gamma_{ij}$ without overbar) is
\be
J = \frac{\sqrt{^{(2)}\gamma}}
         {\sqrt{\gamma} \sqrt{\gamma^{\bar{i}\bar{i}}}} .
\ee
If we insert the latter into the right hand side of Eq.~(\ref{n_i_dAbar})
we find
\be
n_i d\bar{A} = \frac{1}{\sqrt{\gamma}}
               \frac{1}{\sqrt{\gamma^{\bar{3}\bar{3}}}}
               \frac{\partial x^{\bar{3}}}{\partial x^i}
               \sqrt{^{(2)}\gamma} dx^{\bar{1}} dx^{\bar{2}}
             = \frac{1}{\sqrt{\gamma}} N_i dA ,
\ee
where $N_i$ is normalized with the physical metric and $dA$ is the physical
surface element. Let us now define
\be
F^i := \frac{f^i}{\sqrt{\gamma}} .
\ee
Then the right hand side of Eq.~(\ref{Gauss_law_flat}) can be written as
\be
\oint f^i n_i d\bar{A} = \oint F^i N_i dA ,
\ee
while the left hand side is
\be
\int \partial_i f^i d^3x
=
\int \frac{1}{\sqrt{\gamma}}\partial_i (\sqrt{\gamma} F^i) \sqrt{\gamma}d^3x
=
\int D_i F^i \sqrt{\gamma}d^3x ,
\ee
where $D_i$ is the covariant derivative operator. Together this yields
\be
\label{Gauss_law}
\int D_i F^i \sqrt{\gamma}d^3x = \oint F^i N_i dA ,
\ee
which is the well known coordinate independent form of Gau{\ss}'s theorem.

This shows that we can use other metrics besides the physical one in
Gau{\ss}'s theorem, because all pieces of any metric cancel. Yet, whatever
metric we choose to use, must also be used to normalize our normal vector.

\section{On the influence of different normalizations}
\label{normalization-example}

We now discuss the differences between using the physical metric
$\gamma_{ij}$ (as in~\cite{Teukolsky:2015ega,Deppe:2021bhi})
and the flat metric $\delta_{ij}$ to normalize the vectors
$n_i$ normal to a domain boundary.

To obtain a simple example we start with a 2-dimensional spacetime metric
$ds^2 = -\alpha^2 dt^2 + \gamma_{xx} dx^2$.
If we retrace the steps that lead from Eq.~(\ref{Cons0}) to
Eq.~(\ref{Cons2}), we find
\be
\label{Cons2_1d}
\partial_t u +  \partial_x f = s .
\ee
The main step in the DG method consists of integrating the $\partial_x f$
term, which in one spatial dimension becomes
\ba
\int_a^b dx \psi \partial_x f = [\psi f]_a^b - \int_a^b dx f \partial_x\psi
& \rightarrow &
[\psi f^*]_a^b - \int_a^b dx f \partial_x\psi \nonumber \\
& = &
[\psi (f^* - f)]_a^b + \int_a^b dx \psi \partial_x f ,
\ea
where again we have introduced a numerical flux $f^*$. The term
$[\psi (f^* - f)]_a^b$ corresponds to the surface integral in
Eq.~(\ref{introduce_numflux}), and can be written as
\be
\label{boundary-terms}
[\psi (f^* - f)]_a^b = \psi (f^* - f)n|_b + \psi (f^* - f)n|_a ,
\ee
where the outward pointing normals are $n|_b = 1$ and $n|_a = -1$.
So far the physical metric $\gamma_{xx}$ has not appeared. Following
Teukolsky~\cite{Teukolsky:2015ega} we now define a normal vector
$N := n/\sqrt{\gamma^{xx}}$ that is normalized with respect to the
physical metric. We then obtain
\be
\label{fN-phys}
\psi (f^* - f) n = \psi (f^* - f) N \sqrt{\gamma^{xx}} .
\ee
This means we can replace the $n$ that was normalized with respect to the
flat metric with an $N$ that is normalized with respect to the physical metric
$\gamma_{xx}$, provided we include other metric factors. Notice that the
factor $\sqrt{\gamma^{xx}}$ in Eq.~(\ref{fN-phys}) is equivalent to the
$\gamma^{ij}$ under the root in Eq.~(35) of~\cite{Deppe:2021bhi},
and that in the case discussed here $\xi \propto x$, so that the $J$ and
$\partial\xi^{\hat{i}}/\partial x^j$ terms in 
Eq.~(35) of~\cite{Deppe:2021bhi} drop out.
The fact that the $N$ and $\sqrt{\gamma^{xx}}$ terms in Eq.~(\ref{fN-phys})
cancel each other, agrees with the discussion in appendix A
of~\cite{Teukolsky:2015ega} that calls the appearance of the physical metric
illusory, and also with our \ref{Gauss_theorem}.

The situation becomes less trivial when one considers how the numerical flux
$f^*$ is actually computed, which is related to the point
about metric discontinuities being tricky, that is raised
in~\cite{Deppe:2021bhi}.
As an example, let us consider the LLF flux of Eq.~(\ref{LLF_flux}). It
depends on the field value $u_\mathrm{in}$ in the current domain and the
$u_\mathrm{adj}$ from the adjacent domain.
For $n_i$ Eq.~(\ref{LLF_flux}) makes no such distinction because $n_i$,
normalized with the flat metric, is the same on
both sides. The analog to Eq.~(\ref{LLF_flux}) found in Eq.~(36)
of~\cite{Deppe:2021bhi} is
\be
\label{LLF_flux_Deppe}
(f^i N_i)^* = \frac{1}{2}\left[
  f^i(u_\mathrm{in}) N_i^{\mathrm{in}} + f^i(u_\mathrm{adj}) N_i^{\mathrm{adj}}
   + |\Lambda|_\mathrm{max}\left(u_\mathrm{in} - u_\mathrm{adj}\right)\right] ,
\ee
where $N_i^{\mathrm{in}}$ and $N_i^{\mathrm{adj}}$ are the normals in the
different domains that differ if the physical metric is discontinuous
across the domain boundary. Also note that $|\Lambda|_\mathrm{max}$
denotes the absolute maximum eigenvalue magnitude, when we consider
eigenvalues from both sides. I.e. the numerical flux in~\cite{Deppe:2021bhi}
differs from our approach if the physical metric is discontinuous
across the domain boundary. Note, however, that the physical metric of the
true continuum solution will always be continuous, so that such
discontinuities will converge to zero with increasing resolution.

Finally, we will compute the numerical flux with both Eqs.~(\ref{LLF_flux})
and (\ref{LLF_flux_Deppe}) for a simple example.
Consider the case where we have a single field $u$ with $f=u$ and $s=0$,
so that Eq.~(\ref{Cons2_1d}) becomes
\be
\label{Adv_1d}
\partial_t u +  \partial_x u = 0 ,
\ee
which is a simple advection equation for $u$. We wish to solve this
equation in a 1-dimensional domain $x\in [a,b]$.
If we use $n$ as normal, the eigenvalue $\lambda=1$ on the right boundary
(at $x=b$). At $x=b$ Eq.~(\ref{LLF_flux}) then yields
\be
(f n)^* = u_\mathrm{in} = f n.
\ee
If we use $N$ as normal, the eigenvalue $\Lambda=N$ on the right boundary
(at $x=b$).
Thus Eq.~(\ref{LLF_flux_Deppe}) results in
\ba
(f N)^*
&=& \frac{1}{2}\left[
  u_\mathrm{in} N^{\mathrm{in}} + u_\mathrm{adj} N^{\mathrm{adj}}
   + |\Lambda|_\mathrm{max}
     \left(u_\mathrm{in} - u_\mathrm{adj}\right)\right] \nonumber \\
&=& \frac{1}{2}\left[
  (N^{\mathrm{in}}  + |\Lambda|_\mathrm{max}) u_\mathrm{in}
 +(N^{\mathrm{adj}} -|\Lambda|_\mathrm{max}) u_\mathrm{adj}  \right] ,
\ea
where $|\Lambda|_\mathrm{max} = \max(|N^{\mathrm{in}}|,|N^{\mathrm{adj}}|)$.
Unless $N^{\mathrm{in}} = N^{\mathrm{adj}}$, $(f N)^*$ is not equal to
$f N$, and thus the result really differs from $(f n)^* = f n$.
An analogous difference also occurs on the left boundary at $x=a$.

The question now arises which approach we should use. The analytic solution
of the advection Eq.~(\ref{Adv_1d}) is $u=h(x-t)$, where $h(x)$ is an
arbitrary function. I.e. we obtain a profile that moves to
the right over time. Thus no boundary condition is needed on the right,
because nothing is entering the domain from there. The corresponding
numerical flux should thus be computed solely from quantities inside our
domain, and hence be given by the upwind flux
$f^* = f = u_\mathrm{in}$.
The latter is exactly what we have obtained from the LLF flux of
Eq.~(\ref{LLF_flux}), when using the flat metric to normalize our normals.
This is expected, as the LLF flux for a single field obeying
Eq.~(\ref{Adv_1d}) is known to be equivalent to the
upwind flux. Also notice that the boundary term at $x=b$ in
Eq.~(\ref{boundary-terms}) entirely vanishes for this upwind flux, which is
equivalent to not imposing any boundary condition on the right. Yet, we do
not obtain these same results if we follow~\cite{Deppe:2021bhi} and
normalize with the physical metric (unless the physical metric is continuous
across the boundary). Nevertheless, we believe that both normalization
approaches can work, because the physical metric of the true continuum
solution will always be continuous. We thus expect both approaches to
converge to the same physical solution. However, we prefer our approach to
the one in~\cite{Deppe:2021bhi}, because it is simpler, and also because
it reproduces the correct upwind result for a single advection equation.

We should also note, that in the first paper about the SpECTRE
code~\cite{Kidder:2016hev} it is claimed (in the footnote on page 7) that
the ``unit normal'' is the same on the two sides of the boundary. From the
context of this footnote it seems as if the authors mean $N_i$ (normalized
with respect to the physical metric), when they write ``unit normal''. This,
however, cannot be true because it is precisely $n_i$ (normalized with
respect to flat metric), that is the same on both sides of the boundary.
This is because $n_i$ denotes the normal expressed in the global
Cartesian-like $x^i$ coordinates, which cover all domains (see remark after
Eq.~(\ref{dbarA_example})). Thus by definition $n_i$ cannot have any
discontinuities, while $N_i$ (obtained by renormalizing $n_i$ with the
physical metric) can be discontinuous, if the physical metric is.

Another way of seeing that $N_i$ can be discontinuous, is by recalling
that it is normalized with the physical metric and thus
\be
N_i^{\mathrm{in}} N_j^{\mathrm{in}} \gamma^{ij}_{\mathrm{in}} = 1 =
N_i^{\mathrm{adj}} N_j^{\mathrm{adj}} \gamma^{ij}_{\mathrm{adj}} .
\ee
Therefore, if $\gamma^{ij}_{\mathrm{in}}$ and $\gamma^{ij}_{\mathrm{adj}}$
differ, $N_i^{\mathrm{in}}$ and $N_i^{\mathrm{adj}}$ can differ as well.
Also notice, that Eq.~(3.16) of~\cite{Kidder:2016hev} has a term with
eigenvalues, which is identical to the one in Eq.~(\ref{LLF_flux_Deppe}),
and also contains an $N_i$ that is different on both sides of the boundary.
Hence it seems the authors of~\cite{Kidder:2016hev} agree with us, that $N_i$
can be discontinuous.


In Sec.~\ref{GHG_runs} we have tested the evolution of a single black hole
using the DG method, where domain normals are normalized with respect to the
flat metric. As we have seen, the discontinuities in the physical metric are
not a problem for our approach, even though the numerical solution goes through
an initial rapid evolution phase.



\bigskip
\noindent
{\bf References\\}

\bibliographystyle{unsrt}
\bibliography{references}

\end{document}